\documentclass[12pt]{article}
\pdfoutput=1
\usepackage{jheppreprint}
\usepackage{tikz}
\title{Fixed-Topology Lorentzian Triangulations: \newline Quantum Regge Calculus in the Lorentzian Domain}
\author{Kyle Tate \textmd{and} Matt Visser}
\affiliation{School of Mathematics, Statistics, and Operations Research, \\
Victoria University of Wellington, PO Box 600, Wellington 6140, New Zealand}
\emailAdd{kyle.tate@msor.vuw.ac.nz}
\emailAdd{matt.visser@msor.vuw.ac.nz}
\abstract{A key insight used in developing the theory of Causal Dynamical Triangulations (CDTs) is to use the causal (or light-cone) structure of Lorentzian manifolds to restrict the class of geometries appearing in the Quantum Gravity (QG) path integral. By exploiting this structure the models developed in CDTs differ from the analogous models developed in the Euclidean domain, models of (Euclidean) Dynamical Triangulations (DT), and the corresponding Lorentzian results are in many ways more ``physical''. 

In this paper we use this insight to formulate a Lorentzian signature model that is analogous to the Quantum Regge Calculus (QRC) approach to Euclidean Quantum Gravity. We exploit another crucial fact about the structure of Lorentzian manifolds, namely that certain simplices are not constrained by the triangle inequalities present in Euclidean signature. We show that this model is \emph{not} related to QRC by a naive Wick rotation; this serves as another demonstration that the sum over Lorentzian geometries is not simply related to the sum over Euclidean geometries. By removing the triangle inequality constraints, there is more freedom to perform analytical calculations, and in addition numerical simulations are more computationally efficient. 

We first formulate the model in 1+1 dimensions, and derive scaling relations for the pure gravity path integral on the torus using two different measures. It appears relatively easy to generate ``large'' universes, both in spatial and temporal extent. In addition, loop-to-loop amplitudes are discussed, and a transfer matrix is derived. We then also discuss the model in higher dimensions. 

\bigskip
\noindent
24 August 2011; 15 October 2011; \LaTeX-ed \today
}
\begin{document}
\maketitle
\newpage
\section{Introduction} \label{sec: intro}
Prior to the formulation of Causal Dynamical Triangulations (CDTs) in reference \cite{Ambjorn:1998xu}, and further developed in \cite{Loll:2000my, Ambjorn:2002gr, Ambjorn:2005db, Ambjorn:2005qt, Ambjorn:2006jf}, the two main approaches to simplicial Quantum Gravity (QG) were those of   (Euclidean) Dynamical Triangulations (DT) and the (Euclidean) Quantum Regge Calculus (QRC), also known as (Euclidean) Fixed Triangulations (FT). Both of these approaches are formulated in the Euclidean sector of QG, that is, they are sums over geometries which have Euclidean signature. In two dimensions, it is possible to completely solve the DT path integral by using matrix model techniques (see \cite{David:1992jw} and references within, and \cite{Benedetti:2008hc} for the generalization to CDT), however the results are completely unphysical: There exist only two phases of the quantized geometry, a crumpled phase and a polymerized phase \cite{Ambjorn:2005jj}; this rather strongly suggests that DT does not have a physically relevant classical limit. This problem was remedied in CDTs by shifting to a sum over Lorentzian-signature geometries which exhibit a ``causal'' structure. That is, one sums only over that subset of Euclidean geometries that is compatible with the existence of a Lorentzian signature metric.  The key point is that CDTs exploit the causal structure present in Lorentzian geometry, and in so doing lead to a more physically realistic model of Quantum Gravity. (For related comments in continuum quantum gravity, and further background references,  see~\cite{Visser:1989ef}.)

In counterpoint, the QRC approach to Euclidean QG is rather hard to work with analytically \cite{Rocek:1982fr}  (except in the weak field approximation), owing to the fact that each simplex in the triangulation must satisfy generalized triangle inequalities. However, in the Lorentzian domain this difficulty disappears --- completely so in 1+1 dimensions and largely so in higher dimensions --- provided that the simplices in the triangulation are chosen such that there are both time-like and space-like edges. QRC in the Lorentzian domain has been discussed before in reference \cite{Williams:1986hx}, where the inverse free propagator was calculated, however the disappearance of certain triangle inequality constraints was not noted or discussed. 

\enlargethispage{35pt}

In section~\ref{sec: LorTri}, we demonstrate that for Lorentzian triangles with one space-like and two time-like edges (or vice-versa), the relative magnitudes of the edge lengths are completely unconstrained. The Lorentzian triangle inequalities are, in this particular context, vacuous. Exploiting this fact, in section~\ref{sec: 1p1LFT}  we formulate a model of Lorentzian quantum gravity in 1+1 dimensions and demonstrate that it is not simply related to the QRC model in 1+1 dimensions by Wick rotation.  In section~\ref{sec: 1p1analysis}  we derive scaling relations for the pure gravity model on the torus using two different measures, one of them corresponding to the DeWitt measure in 2 dimensions. In section~\ref{sec: loopamp}  we discuss loop-to-loop amplitudes and derive a transfer matrix for the model. In section~\ref{sec: spikes} we will discuss the non-occurrence of spikes in this model. In section~\ref{sec: highdim}  we discuss the configuration space of the theory in higher dimensions. We then conclude with a brief discussion.

\section{Lorentzian Triangles} \label{sec: LorTri}
\begin{figure}[htb]
\centering
\begin{tikzpicture}[scale=0.4]
\draw[gray] (-8,0) -- (8,0) node[anchor = west] {$x$};
\draw[gray] (0,-8) -- (0,8) node[anchor = south] {$t$};
\draw[thick] (-4,-4) -- ( 4,-4);
\filldraw[black] (-4,-4) circle (1pt) node [anchor = north]  {$(-\frac{S}{2},-\frac{S}{2})$} ;
\filldraw[black] (4,-4) circle (1pt) node [anchor = north]  {$(-\frac{S}{2},\frac{S}{2})$} ;
\draw[dashed] (-4,-4) -- (8,8)  (4,-4) -- (-8,8);
\draw[thick] (-4,-4) -- (-2,6)  (4,-4) -- (-2,6);
\filldraw[black] (-2,6) circle (1pt) node [anchor = south] {$(t,x)$};
\draw (0,-4) node [anchor = north west] {$S$};
\draw (-3,1)  node [anchor = south east] {$T_1$}; 
\draw (1,1.45) node[ anchor = south ] {$T_2$};
\end{tikzpicture}
\caption{A Lorentzian triangle with space-like edge length $S$,  and time-like edge lengths $T_1$, $T_2$.}
\label{fig: stt-triangle}
\end{figure}
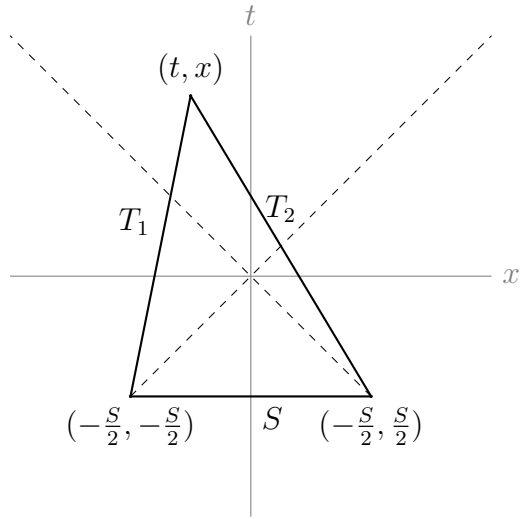

In formulating the Lorentzian-signature Fixed-Topology Triangulation (LFT) model in 1+1 dimensions we consider Lorentzian triangles which have one space-like edge and two time-like edges. Unlike triangles in Euclidean space, these triangles (as well as ones with two space-like and one time-like edge) do \emph{not} have their edge lengths constrained by inequalities. Without loss of generality consider a Lorentzian triangle with one space-like edge length $S$, and two time-like edges of lengths $T_1$ and $T_2$ in the coordinate system shown in figure \ref{fig: stt-triangle}. The third vertex of the triangle lies at the point $(t,x)$, which must be in the future light-cone of both of the other vertices, i.e. $ t >0$ and $|t| \geq |x|$ (the case where it lies in the past light cone follows in an identical manner). Therefore we have
\begin{eqnarray}
T_1^2 &=& t^2 - x^2 + S(t-x) \geq 0,  \label{eq: T12}\\
T_2^2 &=& t^2 - x^2 + S(t+x) \geq 0. \label{eq: T22}
\end{eqnarray}
%

These relations can be inverted to yield
\begin{eqnarray}
t &=& \frac{1}{2S} \left( \sqrt{(S^2 + [T_1 - T_2]^2)(S^2 + [T_1+T_2]^2)} - S^2 \right), \label{eq: tsol} \\
x &=& \frac{1}{2S} \left(T_2^2 - T_1^2\right). \label{eq: xsol}
\end{eqnarray}
Both equations \eqref{eq: tsol} and \eqref{eq: xsol} have a real solution for any value of $S \in (0,\infty)$ and $T_1$, $T_2 \in [0,\infty)$; thus the edge lengths are completely unconstrained. For given values of $S$, $T_1$, and $T_2 $, we note that  in the coordinate system under consideration the third vertex is located at the intersection of the hyperbolae given in equations \eqref{eq: T12} and \eqref{eq: T22}. Using equation \eqref{eq: tsol} we can easily find the area of a Lorentzian triangle in terms of the edge lengths $S$, $T_1$ and $T_2$:

\begin{eqnarray}
A_{L}(S,T_1,T_2) &=& \frac12 \times S \times \left(\frac{S}{2} + t\right), \nonumber \\
                         &=& \frac14 \sqrt{(S^2 + [T_1 - T_2]^2)(S^2 + [T_1+T_2]^2)} ~, \label{eq: triA1}\\
                         &=& \frac14 \sqrt{S^4 + T_1^4 + T_2^4 + 2S^2T_1^2 + 2S^2T_2^2 - 2T_1^2T_2^2} ~.\label{eq: triArea}
\end{eqnarray}
Equation \eqref{eq: triArea} is closely related to, but not identical to, Heron's formula which describes the area of a Euclidean triangle in terms of its edge lengths $a$, $b$, $c$ by:
\begin{equation}
\label{eq: HerArea}
A_{E}(a,b,c) = \frac14 \sqrt{-a^4 -b^4 -c^4 + 2a^2b^2 + 2a^2c^2 + 2b^2c^2} ~.
\end{equation}
The relation between these quantities is what one would expect by a simple Wick rotation, $t \rightarrow it$:
\begin{equation}
\label{eq: Area-Lor-Eu}
A_{L}(S,T_1,T_2) = -i A_{E}(S,iT_1,iT_2),
\end{equation}
where the factor of $-i$ is due to the fact that we should be multiplying Lorentzian area by a factor of $\sqrt{\det(\eta)} = i$. Again, it is clear that these Lorentzian triangles are not constrained by inequalities, for while in equation \eqref{eq: HerArea} the area becomes pure imaginary if the Euclidean triangle inequalities are violated, in equation \eqref{eq: triArea} the area is real for any choice of $S$, $T_1$, $T_2$.

Another quantity that is of interest in studying simplicial gravity is the angle between two edge lengths. For the angle between two space-like edge lengths these are found using the standard trigonometric relations, however finding the angle between a space-like edge and a time-like edge or two time-like edges requires more care. For configurations such as in figure \ref{fig: stt-triangle} these angles will be complex numbers (boosts). A consistent way of assigning angles to these triangles is described in reference~\cite{Sorkin:1975ah}. We denote the angle between $S$ and $T_1$ to be $\theta_1$, between $S$ and $T_2$ to be $\theta_2$, and the angle between $T_1$ and $T_2$ to be $\varphi$. These angles are given by:
\begin{align}
\cos \theta_1 &= \frac{1}{2i} \left( \frac{S^2-T_1^2+T_2^2}{ST_1} \right), &\quad \sin \theta_1 &= \frac{2}{ST_1}~A_L(S,T_1,T_2), \label{eq: ST1angle} \\
\cos \theta_2 &= \frac{1}{2i} \left( \frac{S^2+T_1^2-T_2^2}{ST_2} \right), &\quad \sin \theta_2 &= \frac{2}{ST_2}~A_L(S,T_1,T_2), \label{eq: ST2angle}\\
\cos \varphi &= \frac{1}{2} \left( \frac{S^2+T_1^2+T_2^2}{T_1T_2} \right), &\quad \sin \varphi &= \frac{2}{iT_1T_2}~A_L(S,T_1,T_2). \label{eq: T1T2angle}
\end{align}
There are few things to notice here: First is that $\cos \theta_1$ and $\cos \theta_2$ are purely imaginary, and therefore $\Re[\theta_1] = \Re[\theta_2] = \pi/2$, and second is that $\cos \varphi >1$ so that $\Re[\varphi] = 0$. 

Equipped with these notions of Lorentzian-signature triangles, we shall now define a theory of Lorentzian-signature Quantum Gravity in 1+1 dimensions, which will be closely analogous to QRC. However as we will soon demonstrate, these theories are \emph{not} related by any simple Wick Rotation.

\section{1+1 Lorentzian Fixed Triangulations} \label{sec: 1p1LFT}
Classically two-dimensional Einstein gravity is trivial, since the action is a topological invariant, the Euler characteristic, and is therefore constant with respect to local degrees of freedom in the theory. This is equivalent to the fact that in two dimensions the Einstein tensor vanishes identically. However two-dimensional QG is non-trivial, and due to the topological nature of the Einstein-Hilbert term can often be solved analytically. Two dimensional simplicial QG was the framework within which CDT was first formulated \cite{Ambjorn:1998xu}, and has also been investigated using QRC \cite{Hamber:1985gw, Gross:1990fq, Hamber:1992jh}. Here we will consider pure two dimensional QG, that is, the path integral:
\begin{eqnarray}
Z = \int \mathcal{D}[g] \; e^{i S[g]} ~, \qquad\qquad
S[g] = \Lambda \int d^2x \sqrt{-g} ~. \label{eq: 2DEH}
\end{eqnarray}
In the Regge Calculus \cite{Regge:1961px, Williams:1996jb}, a discrete approach is taken in evaluating equations \eqref{eq: 2DEH}. The manifold is taken to be piecewise flat and is composed, in two dimensions, of triangles glued along shared edges. The metric degrees of freedom are described by the edge-lengths of each triangle, and curvature is localized at the vertices of the triangles. We shall consider a model in which the manifold under consideration either has cylindrical topology $\mathcal{S}^1 \times \mathbb{R}$, where the $\mathcal{S}^1$ are space-like surfaces parameterized by ``time'', or the toroidal topology $\mathcal{S}^1 \times \mathcal{S}^1$.

\subsection{Triangulations} \label{ssec: triangle}

This manifold has a simple triangulation the ``traditional'' version of which (in analogy with common conventions in Euclidean signature) is shown in figure \ref{fig: tricyl}. Here the horizontal edges are taken to be space-like, and the diagonal and vertical lines are taken to be time-like. For the cylinder the time-like boundaries are identified, and for the torus the space-like boundaries are also identified.  A physically equivalent but more ``symmetric'' version of the triangulation is presented in figure \ref{fig: symmetric}. 
\begin{figure}[t]
\centering
\begin{minipage}[b]{7cm}
\begin{tikzpicture}[scale=1.40]
\def \n{2}
\def \s{0.5}
\def \d{1.5}
\draw[step=\s cm,black,thick]
 (-\n,-\n)  grid (\n,\n) ;
 \foreach \x in {-\n,-\d,...,\n}
 	\draw[black,thick] (-\n ,-\x ) -- (\x ,\n ) (-\x,-\n)--(\n,\x);
\foreach \y in {-\n,-\d,...,\n}
	\foreach \z in {-\n,-\d,...,\n}
		\filldraw[black] (\y,\z) circle (0.5pt);
 \end{tikzpicture}
 \end{minipage}
 \begin{minipage}[b]{3.7cm}
 \begin{tikzpicture}[scale=0.75]
\filldraw[black] (0,0) circle (1pt) node[anchor = north] {$(i,j)$};
\draw (0,0) -- (4,0) (0,0) -- (0,4) (0,0) -- (4,4);
\draw (0,2) node[anchor = east] {$\bar{t}_{ij}$};
\draw (2,0) node[anchor = north] {$l_{ij}$};
\draw (2, 2) node[anchor = south east] {$t_{ij}$};
\end{tikzpicture}
\end{minipage}
\caption{An 8$\times$8 triangulation of the cylinder and a vertex in the triangulation with its associated edge lengths. (Traditional version.)}
\label{fig: tricyl}
\end{figure}
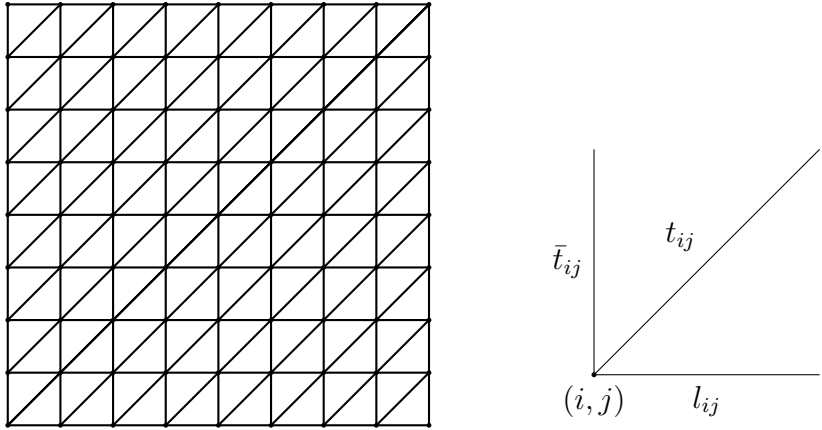
\begin{figure}[htbp]
\begin{center}
\begin{tikzpicture}[scale=1.4]
\def \n{2}
\def \s{0.5}
\def \d{1.5}
\draw[step=1 cm,thick]
 (-2,-2)  rectangle (2,2) ;
\draw[thick] (-2,-1.5)--(2,-1.5);
\draw[thick] (-2,-1)--(2,-1);
\draw[thick] (-2,-0.5)--(2,-0.5);
\draw[thick] (-2,0)--(2,0);
\draw[thick] (-2,0.5)--(2,0.5);
\draw[thick] (-2,1)--(2,1);
\draw[thick] (-2,1.5)--(2,1.5);
 \foreach \x in {-2,-1,...,1}
 	\foreach \u in {-2,-1,...,1}
		\draw[thick] (\u,\x)--(\u+0.5,\x+0.5)--(\u+1,\x);
\foreach \x in {-1.5,-0.5,...,1.5}
 	\foreach \u in {-2,-1,...,1}
		\draw[thick] (\u,\x+0.5)--(\u+0.5,\x)--(\u+1,\x+0.5);
 \end{tikzpicture}
\caption{Regular triangular lattice. (Symmetric version in terms of $S$, $T_1$ and $T_2$.) For each triangle one edge is spacelike, two are timelike (or at worst null). }
\label{fig: symmetric}
\end{center}
\end{figure}
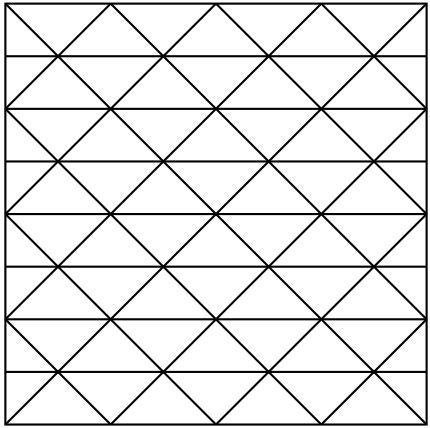
The simplices in either of these triangulations are the Lorentzian-signature triangles described in section~\ref{sec: LorTri}. We take there to be $N_s$ edge-lengths  in the spatial direction, and $N_t$ in the time direction, for a total of $N=2N_sN_t$ triangles in the spacetime.
 Note that by adopting a fixed-topology triangulation of this type one is automatically excluding, at the kinematical level, the possibility of a polymerized phase. By this we mean that the polymer-like phase found in DT \cite{Ambjorn:2005jj} is not realizable in this model, however this does not exclude the possibility of a ``rough'' phase, (that is, a ``fractal'' or ``crumpled'' phase), similar to that found in QRC \cite{Hamber:1985gw, Gross:1990fq, Hamber:1992jh}.  Before we rewrite equations \eqref{eq: 2DEH}  in their simplicial form we point out a few differences between two dimensional Euclidean and Lorentzian Regge calculus.

\subsection{Metric: traditional} \label{ssec: metric}

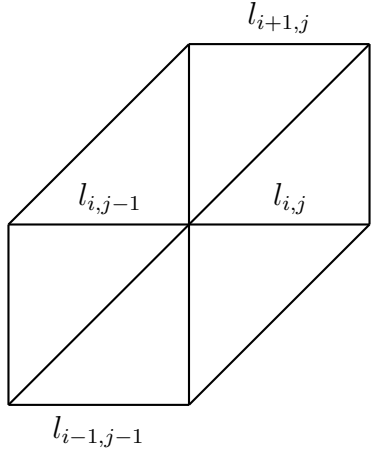
\begin{figure}[!htbp]
\begin{center}
\begin{tikzpicture}[scale=1.2]
\def \n{2}
\def \s{0.5}
\def \d{1.5}
\draw[thick] (-2,0)--(2,0);
\draw[thick] (0,-2)--(0,2);
\draw[thick] (-2,-2)--(2,2);
\draw[thick] (0,2)--(2,2);
\draw[thick] (-2,-2)--(0,-2);
\draw[thick] (-2,-2)--(-2,0);
\draw[thick] (2,0)--(2,2);
\draw[thick] (0,-2)--(2,0);
\draw[thick] (-2,0)--(0,2);
\draw (1.125,0) node[anchor = south] {$l_{i,j}$};
\draw (1,2) node[anchor = south] {$l_{i+1,j}$};
\draw (-0.875,0) node[anchor = south] {$l_{i,j-1}$};
\draw (-1,-2) node[anchor = north] {$l_{i-1,j-1}$};
\end{tikzpicture}
\caption{Traditional triangular lattice: Immediate neighborhood of a specific vertex in the limit where all the timelike edges become null. 
Only the spacelike lengths are labelled. }
\label{fig: traditional-vertex}
\end{center}
\end{figure}

For Euclidean triangulations, the flat metric inside a triangle with edge lengths $l_1$, $l_2$, and $l_3$ (in a coordinate system where $l_1$ is the unit vector in the $1$ direction and $l_2$ is the unit vector in the $2$ direction) is given by:
\begin{equation}
g_{\mu \nu} =  \begin{pmatrix} l_{1}^2 & \frac{1}{2} \{ l_{1}^2 + l_{2}^2 - l_{3}^2\} \\\  \frac{1}{2} \{ l_{1}^2 + l_{2}^2 - l_{3}^2\} & l_{2}^2 \\ \end{pmatrix}. \label{eq: eucmet}
\end{equation}
Note that this satisfies $\det(g) = 4A_E^2(l_1,l_2,l_3)$.
For Lorentzian triangulations, the analogous formula for a triangle with edge lengths $\bar t$, $t$ and $l$ in a ``traditional'' coordinate system where $\bar t$ is the unit vector in the 0 direction and $l$ is the unit vector in the 1 direction is:
\begin{equation}
g_{\mu \nu} =  \begin{pmatrix} -{\bar t}^2 & \frac{ 1}{2}\{l^2 - \bar t^2 + t^2\} \\  \frac{ 1}{2} \{l^2 - \bar t^2 + t^2\} & l^2 \\ \end{pmatrix}. \label{eq: lormet}
\end{equation}
Note that this satisfies $\det(g) = -4A_L^2(l,\bar t, t)$.
Again, we note that the metric in equation \eqref{eq: eucmet} switches signature if the triangle inequalities are violated, however equation \eqref{eq: lormet} remains in Lorentzian signature for all real values of $\bar t$, $t$, $l$. In Euclidean Regge calculus the deficit angle at a vertex is given by $\delta = 2\pi - \sum_{s} \theta_s$, where the sum is over the interior angles $\theta_s$ of each triangle $s$ incident on that vertex. In this case the deficit angle is always a real number. By considering equations \eqref{eq: ST1angle}--\eqref{eq: T1T2angle} we see that for Lorentzian triangulations such as in figure \ref{fig: tricyl} the angles summed around any vertex will be $2\pi - i \eta$. So now the deficit will be $\delta = i \eta$ which is purely imaginary. Therefore curvature will be described by a pure boost, and this will be important in developing higher derivative 1+1 theories which we discuss in the conclusions. For any specific vertex $(i,j)$ the deficit angle is in general a complicated function of 12 edge lengths. However, in the limit in which all time-like edge lengths become null, (see figure \ref{fig: traditional-vertex} and section \ref{sec: CausDiam}), the deficit angle simplifies to \cite{Tate:2011aa}:
\begin{equation}
\label{eq: causdiamdeficit}
\delta = i \eta = i \ln \left( \frac{l_{i,j}^2 \; l_{i,j-1}^2}{l_{i+1,j}^2\; l_{i-1,j-1}^2} \right).
\end{equation}
This limit is discussed in more detail in section \ref{sec: CausDiam}.

\subsection{Metric: symmetric} \label{ssec: metric2}

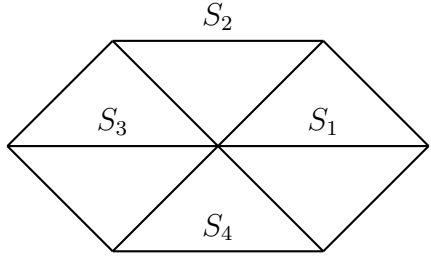
\begin{figure}[htbp]
\begin{center}
\begin{tikzpicture}[scale=1.4]
\def \n{2}
\def \s{0.5}
\def \d{1.5}
\draw[thick] (-2,0)--(2,0);
\draw[thick] (-1,1)--(1,1);
\draw[thick] (-1,-1)--(1,-1);
\draw[thick] (-1,1)--(1,-1);
\draw[thick] (1,1)--(-1,-1);
\draw[thick] (-1,1)--(-2,0);
\draw[thick] (-1,-1)--(-2,0);
\draw[thick] (1,1)--(2,0);
\draw[thick] (1,-1)--(2,0);
\draw (1,0) node[anchor = south] {$S_1$};
\draw (0,1) node[anchor = south] {$S_2$};
\draw (-1,0) node[anchor = south] {$S_3$};
\draw (0,-1) node[anchor = south] {$S_4$};
\end{tikzpicture}
\caption{Regular triangular lattice: Immediate neighborhood of a specific vertex in the limit where all the timelike edges become null.
Only the spacelike lengths are labelled.}
\label{fig: symmetric-vertex}
\end{center}
\end{figure}

In terms of the ``symmetric'' version of the triangulation one proceeds as follows: As in section \ref{sec: LorTri}, choose an initial coordinate system in which the vertices at the ends of the space-like edge of length $S$ lie at $(-S/2,-S/2)$ and $(-S/2,S/2)$ respectively. The third vertex is assumed to be in the forward lightcone of these two vertices at a point $(t,x)$.  For this calculation the key result we need is:
\begin{eqnarray}
-(t+S/2)^2+x^2 &=& -\frac12 \left( \frac12 S^2 +T_1^2+T_2^2 \right).
\end{eqnarray}
Now choose the time-like basis covector to be $\widehat{E}^0{}_\mu =\eta_{\mu \nu} \; \frac12 (\vec{T}_1+\vec{T}_2 )^{\nu} = (-(t+S/2),x)$, and the space-like basis covector to be 
$\widehat{E}^1{}_\mu = (0, S/2)$. The (flat) metric inside the triangle is given by:
\begin{equation}
g_{\mu \nu} = \eta_{\alpha \beta} \; \widehat{E}^{\alpha}{}_{\mu} \widehat{E}^{\beta} {}_{\nu}
= -\widehat{E}^{0}{}_{\mu} \widehat{E}^{0}{}_{\nu} + \widehat{E}^{1}{}_{\mu} \widehat{E}^{1}{}_{\nu}. 
\end{equation}
Calculating these out directly gives:
\begin{align}
g_{00} &= -(t+S/2)^2 + x^2 = -\frac12 \left( \frac12 S^2 +T_1^2+T_2^2 \right),
 \\
g_{01} &= g_{10} = \frac12 x S = \frac14 (T_2^2 - T_1^2),
 \\
g_{11} &= \frac14 S^2.
\end{align}
So the metric in this ``symmetric'' coordinate system is:
\begin{equation}
g_{\mu \nu} = \frac12 \begin{pmatrix} -\{\frac12S^2+T_1^2+T_2^2\} & \;\;\frac{ 1}{2}\{  -T_1^2 + T_2^2\} \\  \frac{ 1}{2} \{- T_1^2 + T_2^2\} & \frac12 S^2 \\ \end{pmatrix}.
\label{eq: g-symmetric}
\end{equation}
Note that this satisfies $\det(g) = -A_L^2(S,T_1,T_2)$. In general the deficit  angle is quite complicated, however  in the limit in which all time-like edge lengths become null, (see figure \ref{fig: symmetric-vertex}, and the discussion in section \ref{sec: CausDiam}), the deficit angle simplifies to \cite{Tate:2011aa}:
\begin{equation}
\label{eq: causdiamdeficit2}
\delta = i \eta = i \ln \left( \frac{S_1^2 \; S_3^2}{S_2^2\; S_4^2} \right).
\end{equation}
%

\subsection{Action} \label{ssec: action}

We can rewrite equations \eqref{eq: 2DEH} in their simplicial form as:
\begin{eqnarray}
Z &=& \int \mathcal{D}[g] \; e^{i S[g]} \rightarrow Z_{LFT}=\int_{\{l\}} d\mu[\{l^2\}] \; e^{i S_{L}[\{l\}]} ~,\label{eq: ZLFT} \\
S &=&  \Lambda \int d^2x \sqrt{-g} \rightarrow S_{L}=\lambda \sum_{s} A_{L}(\{l_{s}\}), \label{eq:SLFT}
\end{eqnarray}
where $\{l\}$ denotes the set of all edge lengths in the triangulation, and $s$ is an index for each triangle in the triangulation. Furthermore $\{l_{s}\}$ is the set of edge lengths for the triangle $s$ (we will denote these $\{l_s,t_s, \bar{t}_s\}$) and $d\mu[\{l\}]$ is the measure on the set of edge lengths which we discuss later in the paper. It is instructive to compare equation \eqref{eq: ZLFT} to the Euclidean path integral of QRC:
\begin{eqnarray}
Z_{QRC} &=& \int_{\{l\}} d\mu[\{l\}] \; \chi[\{l\}] \; e^{- S_{E}[\{l\}]} \label{eq: ZQRC}~, \\
S_{E} &=& \lambda \sum_{s} A_{E}(\{l_{s}\}), \label{eq:SQRC}
\end{eqnarray}
where $\chi[\{l\}]$ is an ``indicator'' function on the set of edge lengths which is $1$ when every edge length satisfies the triangle inequalities and $0$ otherwise. We can Wick rotate equation \eqref{eq: ZLFT}, $t \rightarrow it$, to Euclidean signature to obtain:
\begin{eqnarray}
 Z_{LFT}\rightarrow Z_{LFT,Wick}&=&\int_{\{l\}_{\triangle}} d\mu[\{l\}_{\triangle}] \; e^{-S_E[\{l\}_{\triangle}]}  
 + \int_{\{l\}_{\triangle\!\!\!\!/}} d\mu[\{l\}_{\triangle\!\!\!\!/}] \; e^{-S_E[\{l\}_{\triangle\!\!\!\!/}]}~, \nonumber \\
 &=& Z_{QRC} +  \int_{\{l\}_{\triangle\!\!\!\!/}} d\mu[\{l\}_{\triangle\!\!\!\!/}] \; e^{-S_E[\{l\}_{\triangle\!\!\!\!/}]}~. \label{eq: QRCLFTrelate}
\end{eqnarray}
Here $\{l\}_{\triangle}$ denotes the set of edge-length configurations in the Lorentzian theory which satisfy the Euclidean triangle inequalities, and $\{l\}_{\triangle\!\!\!\!/}$ denotes the set of configurations which explicitly violate them (possibly only one edge). $S_E[\{l\}_{\triangle\!\!\!\!/}]$ is a pure imaginary number, and thus the Wick rotated Lorentzian path integral differs from the Euclidean path integral of QRC by
\begin{equation}
\label{eq: QRCLFTdiff}
Z_{LFT,Wick} - Z_{QRC} =  \int_{\{l\}_{\triangle\!\!\!\!/}} d\mu[\{l\}_{\triangle\!\!\!\!/}] \; e^{-S_E[\{l\}_{\triangle\!\!\!\!/}]}~.
\end{equation}
The fact that Euclidean and Lorentzian quantum gravity are not related by a simple Wick rotation has been known for over a decade (see for example \cite{Ambjorn:1998xu} and \cite{Loll:2000my}). Equation \eqref{eq: QRCLFTdiff} explicitly demonstrates that in this approach finding the relation between the two is at least as hard as fully solving the theory itself.  

Taking the $Z_{LFT}$ of equation \eqref{eq: ZLFT} as primary, we still need some method akin to Wick rotation so that the integral in equation \eqref{eq: ZLFT} converges. The way we do this is to Wick rotate the lattice cosmological constant: $\lambda \rightarrow i \lambda$ to give
\begin{equation}
Z_{LFT,E} \rightarrow \int_{\{l\}} d\mu[\{l\}] \; e^{- S_{L}[\{l\}]} ~.\label{eq: Zwick}
\end{equation}
We note that this is analogous to the method used in CDT \cite{Ambjorn:1998xu}; all results in the end will have to be analytically continued back in $\lambda$. 
Note that with this prescription in Euclidean signature one still integrates over configurations satisfying the (vacuous) Lorentzian triangle inequalities weighted by the Lorentzian area functional. 
Analyzing equation \eqref{eq: Zwick} will be the focus of the next section.

\section{Analysis of the 1+1 Path Integral} \label{sec: 1p1analysis}
In QRC, the issue of which measure on edge lengths, $d\mu[\{l\}]$, to use has been the subject of some contention \cite{Hamber:1997ut, Ambjorn:1997ub}. In this paper we will follow reference \cite{Hamber:1997ut} and will consider measures of the form
\begin{equation}
\label{eq: genmeas}
d\mu[\{l\}] = \prod_{s}[V_d(s)]^{\sigma} \prod_{ij} dl_{ij}^2. 
\end{equation}
Here $V_d(s)$ is the volume of the $d$-dimensional simplex $s$, (for the time being $d=2$), and $\sigma$ is a parameter of the theory related to quantum gravity in the continuum by
\begin{equation}
\sigma = {(d+1)[(1-\omega)d-4]\over 4},
\end{equation}
where $\omega$ is determined by the norm chosen in the superspace of metrics:
\begin{equation}
\label{eq: contnorm}
|| \delta g ||^2 = \int d^4x~ (g(x))^{\omega/2} ~G^{\mu \nu , \alpha \beta}[g(x);\omega]~\delta g_{\mu \nu} \; \delta g_{\alpha \beta}.
\end{equation}
The case $\omega = 0$ is called the DeWitt measure, and in $d=4$ corresponds to $\sigma = 0$. However, in $d=2$ we have $\sigma = -3/2$. There is no unique preferred way to choose $\omega$, therefore in this paper we consider two specific measures: 
\begin{enumerate}
\item the simplest possible case --- the uniform measure $\sigma = 0$;
\item the DeWitt measure $\sigma = -3/2$. 
\end{enumerate}

\subsection{1+1 Lorentzian Fixed Triangulations with Uniform Measure} \label{sec: sigzero}
Consider a triangulation of the torus which has $N$ triangles and $n$ edge lengths arranged as in figure \ref{fig: tricyl}. We wish to investigate equation \eqref{eq: Zwick} with the measure given in equation \eqref{eq: genmeas} for $\sigma = 0$:
\begin{equation}
\label{eq: Zsigzero}
Z = \int \prod_{i=1}^n dl_i^2 ~e^{-\lambda \sum_{j=1}^N A_j},
\end{equation}
where $A_j$ is the area of the $j^\text{th}$ triangle in the triangulation. (Note that $2n=3N=6 N_s N_t$, so $n = 3N_s N_t$.) Consider the Lorentzian area function for one triangle with space-like edge length $l^2 = z$, and time-like edge lengths $t^2 = x$ and $\bar{t}^2 = y$:
\begin{equation}
\label{eq: Acart}
A(z,x,y) = \frac14 \sqrt{z^2 + x^2 + y^2 + 2zx + 2zy - 2xy}.
\end{equation}
This can be rewritten, by expressing $(x,y,z)$ in standard spherical polar coordinates, as
\begin{equation}
\label{eq: Aspher}
A(r,\theta,\phi) = \frac r4 \; \sqrt{1 + \sin 2\theta (\cos \phi + \sin \phi) - \sin^2\theta \sin 2 \phi} = \frac r4 \; f(\theta,\phi),
\end{equation}
which has zeros at $r=0$ and $(r,\theta,\phi)=(r,\pi/2,\pi/4)$. The limits of integration for time-like edges are $[0,\infty)$, however the integral diverges if we allow the space-like edges to go to $0$. This makes perfect sense, as the triangles remain well behaved when the time-like edges are $0$, i.e. in the limit in which they become null. However, when the space-like edge goes to $0$, the triangle becomes degenerate. Therefore we need to regulate the integration in such a way that the degenerate configurations are avoided. The simplest way to do this would be to place a lower limit on the space-like edge lengths; however, for our purposes it is more convenient to remove a thin wedge from the single triangle configuration space by only allowing the azimuthal angle $\theta$ to take values in $[0,\frac{\pi}{2} - \epsilon]$. We can make use of equation \eqref{eq: Aspher} by changing coordinates in the configuration space to $n$-dimensional hyperspherical coordinates, where $r$ is now determined by the edge lengths of all the triangles, $s$, by $r^2 = \sum_s x_s^2 + y_s^2 + z_s^2$. Equation \eqref{eq: Zsigzero} becomes:
\begin{equation}
\label{eq: Zsigzerospher}
Z = \int d\Omega_{n-1} \int_{0}^{\infty} dr~r^{n-1}\; e^{-\frac{\lambda}{4} F(\Omega_{n-1})r} .
\end{equation}
Here $F(\Omega_{n-1})$ is a complicated function of the $n-1$ hyperspherical angles which has a zero only when all of the angles are chosen such that the area of every triangle is zero; these can be avoided provided we apply a regulation to the hyperspherical coordinates that is analogous to the one described above. For the uniform measure, the exact form of $F(\Omega_{n-1})$ is unimportant in determining scaling relations since we can write:
\begin{equation}
\label{eq: Zsigzerosolve} 
Z = \int d\Omega_{n-1} \; \Gamma(n) \; \left(\frac{4}{\lambda \, F(\Omega_{n-1})} \right)^n = \alpha(\epsilon)\; \frac{8^N \, \Gamma(3N/2)}{\lambda^{3N/2}},
\end{equation}
where we have used the topological fact $2n=3N$ (each of a triangle's 3 edges is shared by 2 triangles). We have introduced $\alpha(\epsilon)$ which is just a numerical factor depending on the UV cutoff $\epsilon$ given by:
\begin{equation}
\label{eq: alphasigzero}
\alpha(\epsilon)= \int_{S_{\epsilon}} d\Omega_{n-1} \; \frac{1}{F(\Omega_{n-1})^{3N/2}}~,
\end{equation}
where the zero of $F(\Omega_{n-1})$ is avoided by only integrating over the $\epsilon$ regulated surface $S_{\epsilon}$. The numerical constant is unimportant in calculating how the average area of a triangle scales with $\lambda$:
\begin{equation}
\label{eq: averAsigzero}
\frac{\langle A \rangle}{N} = -\frac{1}{N} \; \frac{d \ln(Z)}{d \lambda} = \frac32 \;\frac1{\lambda}
\end{equation}
This scaling relation for the area is identical to that found in QRC \cite{Bittner:1999tx}, this is essentially because in both path integrals, every edge length can be rescaled by $l^2 \rightarrow l^2/\lambda$. Thus, for the uniform measure, one must look beyond the average area scaling relation to find differences between the two theories.

Note also that in view of the fixed topology, (of the triangulated manifold), and the Gauss-Bonnet theorem, we have
\begin{equation}
\label{eq: Rvert}
\sum_{v}\delta_v = 0.
\end{equation}
This is a sum over all vertices $v$, and is guaranteed to vanish at the kinematical level for each individual geometry in the path integral. (It is very easy to verify this result explicitly for the limit where all timelike edges become null.) This implies an exact result for the spacetime average value of the Ricci scalar which holds for any choice of the measure:
\begin{equation}
\label{eq: averR}
\langle R \rangle = 0.
\end{equation}

\subsubsection{Causal Diamond Truncated Model} \label{sec: CausDiam}

We now discuss a simplifying truncation of the model which will also be of use in the next section, but is worth considering here: that is the limit in which all time-like edges become null. In this case the triangulation is made up of ``causal diamonds'' formed by the two triangles connected by each space-like slice; these causal diamonds have area:
\begin{equation}
\label{eq: diamond area}
A(s) = \frac14 s^2,
\end{equation}
With reference to equation \eqref{eq: g-symmetric}, the metric inside each causal diamond becomes
\begin{equation}
\label{eq: g-diamond}
g_{\mu \nu} =  \frac{s^2}{4} \begin{pmatrix} -1 & ~0 \\  ~0 & ~1 \\ \end{pmatrix} = \frac{s^2}{4} \; \eta_{\mu \nu},
\end{equation}
where $s$ is acting like a conformal factor. With this truncation the integration in equation \eqref{eq: Zsigzero} is trivial:
\begin{equation}
\label{eq: diamondint}
Z = \int \prod_{i} d(s_i^2)\;  e^{-\frac12 \lambda \sum_{i} s_i^2} = \left( \int_0^\infty dx \; e^{-\frac12 \lambda x} \right)^{N/2} = \frac{2^{N/2}}{\lambda^{N/2}}.
\end{equation}
(There are $N/2=N_s\,N_t$ spacelike edges.)
This yields  the simplified scaling relation
\begin{equation}
\label{eq: diamondarea}
\frac{\langle A\rangle }{N} = \frac{1}{2\lambda}.
\end{equation}
It is also trivial to find the average length of the spatial slices in this truncation:
\begin{equation}
\langle L \rangle = N_s \langle s \rangle =   N_s \left( \int_0^\infty dx \, x^{1/2} \; e^{-\frac12 \lambda x} \right) \left( \int_0^\infty dx \; e^{-\frac12 \lambda x} \right)^{-1}= N_s \sqrt{\frac{\pi}{2}} \; \lambda^{-1/2}.
\end{equation}
Note that is is now very easy to generate spatially large universes --- one just has to choose $N_s$  (and $\lambda$) appropriately. Geometrically each of these null-edged triangles extends a distance $s/2$ into the time direction, so the spacetime has temporal extent
\begin{equation}
\langle T \rangle = N_t \; {\langle s \rangle\over2} = N_t \sqrt{\frac{\pi}{2}} \; {\lambda^{-1/2}\over2}.
\end{equation}
So it is also very easy to generate long-lived universes --- one just has to choose $N_t$  (and $\lambda$) appropriately. 
Now observe
\begin{equation}
{\langle A\rangle\over\langle L\rangle\; \langle T \rangle } = {\langle s^2 \rangle\over \langle s \rangle^2}= {4\over\pi} \approx 1.27...
\end{equation}
This can be interpreted as saying that the ``typical'' geometry is not too crumpled, and suggests we are dealing with a relatively ``smooth'' phase.

\subsection{1+1 Lorentzian Fixed Triangulations with the DeWitt Measure} \label{sec: sigDeWitt}
The analysis in this section proceeds in an identical way to the previous section, except now with a non-trivial measure, specifically $\sigma = -\frac32$. The path integral is now:
\begin{equation}
\label{eq: ZDeWitt}
Z = \int \prod_{j=1}^{N} [A_j]^{-3/2} \; \prod_{i=1}^n dl_i^2 \; e^{-\lambda \sum_{j=1}^N A_j}.
\end{equation}
We can again switch to hyperspherical coordinates, where now we denote $(\theta_j,\phi_j)$ as the hyperspherical angles for the edge lengths corresponding to the triangle $j$. With this we can write:
\begin{eqnarray}
Z &=& \int \prod_{j=1}^N \left[ \frac{r}{4} f(\theta_j,\phi_j) \right]^{-3/2} r^{n-1} \; e^{-\frac{\lambda}{4}r \sum_{j=1}^{N} f(\theta_j,\phi_j)}~dr d\Omega_{n-1} \nonumber \\
   &=& 8^N \int \frac{d\Omega_{n-1}}{G(\Omega_{n-1})^{3/2}} \int r^{-1+n-3N/2} \; e^{-\frac{\lambda}{4} F(\Omega_{n-1}) r} ~dr, \label{eq: ZDeWittspher}
\end{eqnarray}
where $G(\Omega_{n-1}) = \prod_j f(\theta_j,\phi_j)$ and $F(\Omega_{n-1}) = \sum_{j} f(\theta_j,\phi_j)$. Both these functions are more complicated than just products and sums of functions of the form $f(\theta,\phi)$ in equation \eqref{eq: Aspher}, due to the fact that the same angles will appear multiple times depending on the incidence matrix of the chosen triangulation. We also note that $F(\Omega_{n-1}) \le \sqrt{2}N$ and  $G(\Omega_{n-1}) \le 2^{N/2}$. We again make use of the relation $2n = 3N$ to write:
\begin{eqnarray}
Z &=&  8^N \int \frac{d\Omega_{n-1}}{G(\Omega_{n-1})^{3/2}} \int_{\epsilon}^{\infty} r^{-1} \; e^{-\frac{\lambda}{4} F(\Omega_{n-1}) r} ~dr \nonumber \\
    &=& 8^N \int \frac{d\Omega_{n-1}}{G(\Omega_{n-1})^{3/2}} \; \Gamma \left[0;\frac{\lambda \epsilon}{4} F(\Omega_{n-1})\right]. \label{eq: ZDeWittinter}
\end{eqnarray}
In contrast to equation \eqref{eq: Zsigzerospher} to ensure convergence we have now had to introduce an explicit ultraviolet cutoff in the $r$ integration, as well as the angular regulation introduced in section \ref{sec: sigzero}. This means that the smallest the space-like edge can become is $\epsilon \sin(\epsilon) \sim \epsilon^2$. Here $\Gamma[0;x]$ is the \emph{incomplete Gamma function} which has the series expansion:
\begin{equation}
\label{eq: gammaexpand}
\Gamma[0;x] = -\gamma_E - \ln(x) + \sum_{k=1}^{\infty} \frac{(-1)^{k+1}x^k}{k \, k!}.
\end{equation}
This converges for all $x \in (0,\infty)$. Equation \eqref{eq: ZDeWitt} is then given by:
 \begin{equation}
 \label{eq: ZDeWittsolve}
Z = \alpha(\epsilon) + \beta(\epsilon) \ln \left( \frac{1}{\lambda} \right) + \sum_{k=1}^{\infty} \alpha_k(\epsilon) \; \epsilon^k \lambda^k,
\end{equation}
with
\begin{eqnarray}
\alpha(\epsilon) &=& 8^N \int_{S_{\epsilon}} \frac{d\Omega_{n-1}}{G(\Omega_{n-1})^{3/2}} \left\{ \ln \left( \frac{4}{\epsilon F(\Omega_{n-1})} \right) - \gamma_E \right\}, \label{eq: alphaDeWitt} \\
\beta(\epsilon)   &=& 8^N \int_{S_{\epsilon}} \frac{d\Omega_{n-1}}{G(\Omega_{n-1})^{3/2}} \label{eq: betaDeWitt}~, \\
\alpha_k(\epsilon) &=& \frac{(-1)^{k+1}}{4^k k \, k!}8^N  \int_{S_{\epsilon}} d\Omega_{n-1} \frac{F(\Omega_{n-1})^k}{G(\Omega_{n-1})^{3/2}}. \label{eq: alphakDeWitt}
\end{eqnarray}
Here $\alpha(\epsilon)$, $\beta(\epsilon)$, and $\alpha_k(\epsilon)$ are numerical factors that depend on the ultraviolet  cutoff $\epsilon$. 

In contrast to the situation for uniform measure, the scaling relation for the average area is now non-trivial: 
\begin{equation}
\label{eq: AscaleDeWitt}
\langle A\rangle  = \frac{1}{\lambda} \left( \frac{\beta(\epsilon) - \sum_{k=1}^{\infty} k \alpha_k(\epsilon) \epsilon^k \lambda^k}{\alpha(\epsilon)-\beta(\epsilon)\ln(\lambda) + \sum_{k=1}^{\infty} \alpha_k(\epsilon) \epsilon^k \lambda^k} \right).
\end{equation}
While this relation demonstrates that the case of the DeWitt measure is less trivial than the case of the uniform measure, it does not give much more information than that. It may be possible that for a fixed value of $\epsilon$ there is a value of $\lambda$ where the average area vanishes or diverges, however (without knowing the precise values of the cutoff-dependent quantities  $\alpha$, $\beta$, and $\alpha_k$) determining if this can happen is impossible. Attempts at performing the integrals in equations \eqref{eq: alphaDeWitt}--\eqref{eq: alphakDeWitt} will run into similar  obstacles to those presented in the next section, therefore ultimately a numerical analysis will be necessary for further investigation of the full model.

\subsubsection{Causal Diamond Truncated Model} \label{sec: CausDiam2}

To further investigate equation \eqref{eq: ZDeWitt} we can look at it in the causal diamond truncation outlined in section \ref{sec: CausDiam}. Again, the integration again becomes tractable, but still exhibits behavior similar to equation \eqref{eq: ZDeWittinter}:
\begin{eqnarray}
Z &=& \int \prod_{i} \frac{8}{(s_i^2)^{3/2}} \; d(s_i^2) \; e^{-\frac{\lambda}{2} \sum_{i} s_i^2}, \nonumber \\
    & =& 8^{N/2} \left( \int_{\epsilon}^{\infty} \frac{dx}{x^{3/2}} \; e^{-\frac{\lambda}{2} x} \right)^{N/2},   \nonumber \\
    &=& 8^{N/2} \left(\frac{2 e^{-\frac{\lambda}{2} \epsilon}}{\sqrt{\epsilon}} - \sqrt{2\lambda} \; \Gamma\left[{1\over2}; {\frac{\lambda \epsilon}{2}}\right] \right)^{N/2}.
    \label{eq: diamondDeWitt}
\end{eqnarray}
\begin{figure}[!htbp]
\begin{center}
\begin{minipage}[b]{7cm}
\centerline{\includegraphics[width=0.80\textwidth, height=0.80 \textwidth]{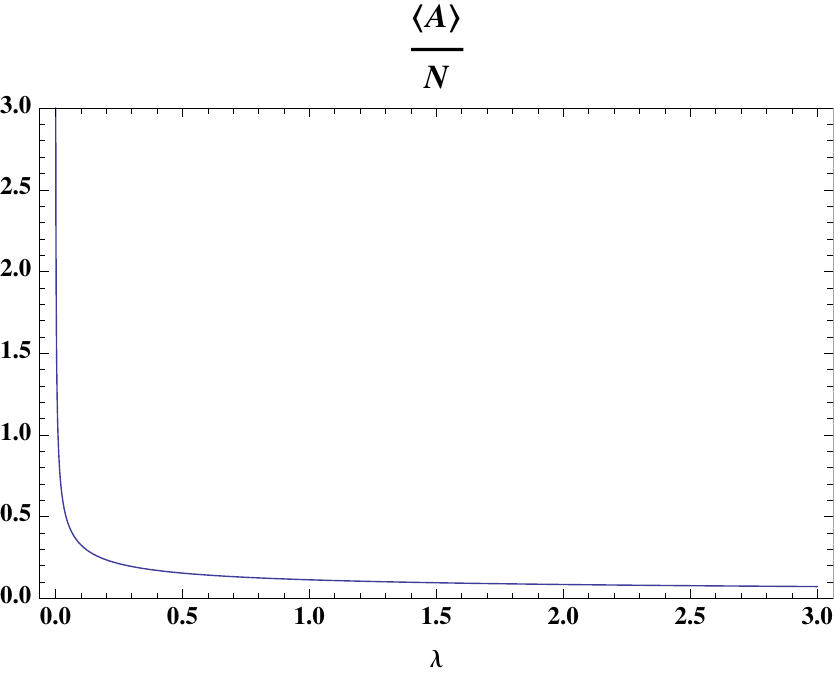}}
\end{minipage}
\begin{minipage}[b]{7cm}
\centerline{\includegraphics[width=0.80\textwidth, height=0.80 \textwidth]{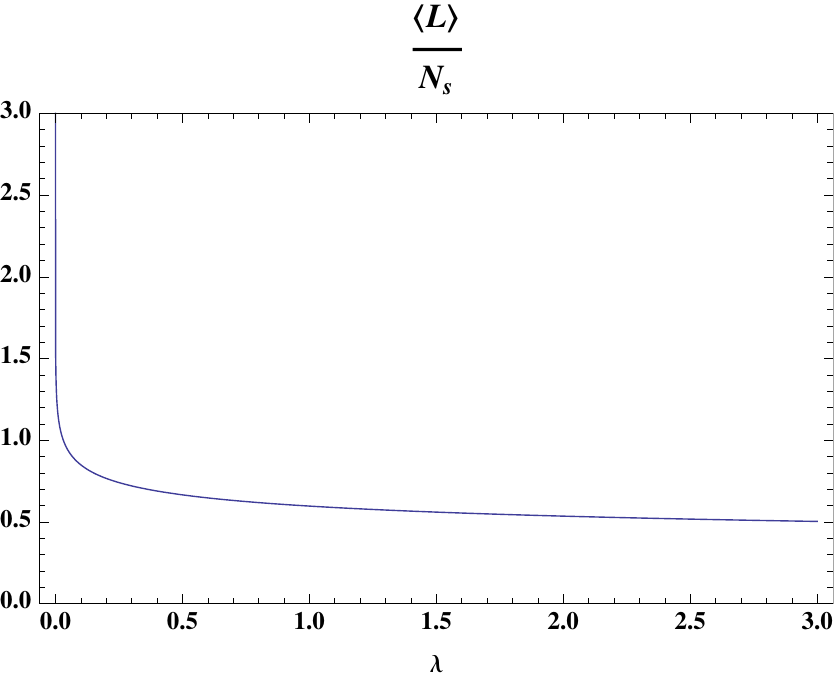}}
\end{minipage}
\caption{A plot of $\frac{\langle A \rangle}{N}= {1\over4}\langle s^2\rangle$ and $\frac{\langle L \rangle}{N_s}= \langle s\rangle$ for the causal diamond truncated model with  DeWitt measure and $\epsilon = 0.1$.}
\label{fig: avgAplot}
\end{center}
\end{figure}

\begin{figure}[!htbp]
\begin{center}
\centerline{\includegraphics[width=0.40\textwidth, height=0.40 \textwidth]{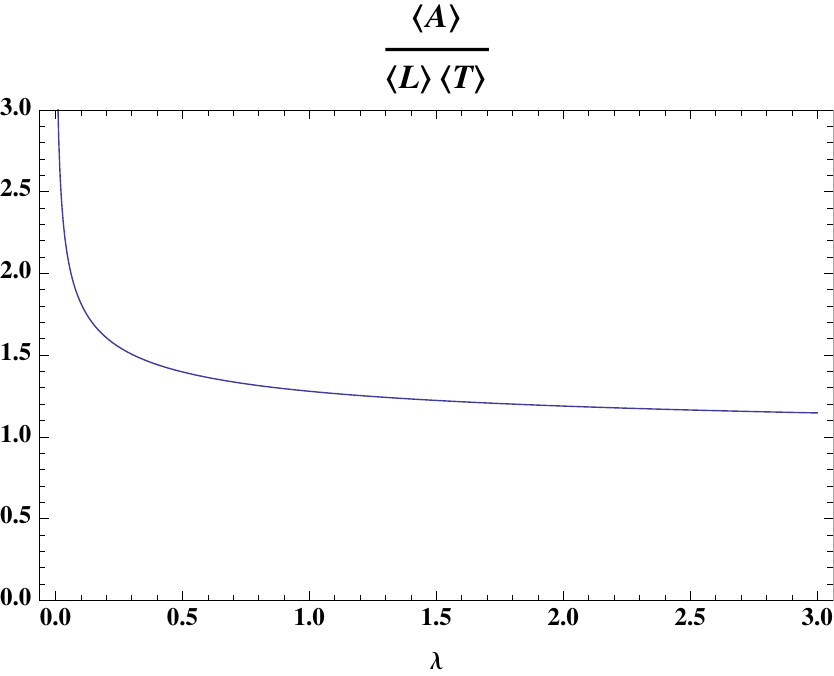}}
\caption{A plot of $\frac{\langle A \rangle}{\langle L\rangle\; \langle T\rangle}= {\langle s^2\rangle\over \langle s\rangle^2}$ for the causal diamond truncated model with  DeWitt measure and $\epsilon = 0.1$.}
\label{fig: avgALTplot}
\end{center}
\end{figure}

The average area then scales as:
\begin{equation}
\label{eq: averageADeWitt}
\frac{\langle A\rangle}{N} 
= \frac{1}{4\lambda} 
{\Gamma\left[{1\over2};{\lambda\epsilon\over2}\right] \over \sqrt{2\over\lambda\epsilon} \; e^{-\lambda\epsilon/2} - \Gamma\left[{1\over2};{\lambda\epsilon\over2}\right] }
= \frac{1}{4\lambda}\,\sqrt{\frac{\pi \epsilon \lambda}{2}} + \mathcal{O}(\epsilon).
\end{equation}
This is a limiting case of equation \eqref{eq: AscaleDeWitt} and is well behaved for all finite $\lambda$ (see figure \ref{fig: avgAplot}), which suggests that equation \eqref{eq: AscaleDeWitt} is as well. The average length of the spatial slices can also be computed by using $\langle L \rangle = N_s \, \langle s \rangle$, while the average temporal extent of the spacetime is $\langle T \rangle = {1\over2} N_t \, \langle s \rangle$, where it is easy to establish that
\begin{equation}
\label{eq: averLdewitt}
\langle s \rangle 
= \frac{ \sqrt{\epsilon}\;\Gamma[0; \frac{\epsilon \lambda}{2}]}{2 e^{- \frac{\epsilon \lambda}{2}} - \sqrt{2\epsilon \lambda}\; 
\Gamma\left[{1\over2}; {\epsilon \lambda \over 2}\right]} 
= \frac{\sqrt{\epsilon}}{2} \, \left(\ln\left(\frac{2}{\epsilon \lambda} \right) - \gamma_E \right)+\mathcal{O}(\epsilon \lambda^{1/2}).
\end{equation}
(See figure \ref{fig: avgAplot}.) In this situation we furthermore have
\begin{equation}
{\langle A\rangle\over\langle L\rangle\; \langle T \rangle } = {\langle s^2 \rangle\over \langle s \rangle^2}= 
{ 
2 \Gamma[{1\over2}; \frac{\epsilon \lambda}{2}] \left(  \sqrt{2\over\lambda\epsilon} \; e^{-\lambda\epsilon/2} - \Gamma\left[{1\over2};{\lambda\epsilon\over2}\right] \right)
\over
\Gamma[0; \frac{\epsilon \lambda}{2}]^2
}.
\end{equation}
(See figure \ref{fig: avgALTplot}.) 
As $\lambda\epsilon\to 0$ this blows up as $\mathcal{O}\left((\lambda\epsilon)^{-1/2} [\gamma_E + \ln(\lambda\epsilon/2)]^{-2}\right)$. In contrast as $\lambda\epsilon\to \infty$ this smoothly tends to unity from above. 
\begin{equation}
{\langle A\rangle\over\langle L\rangle\; \langle T \rangle } = {\langle s^2 \rangle\over \langle s \rangle^2}=
1 + {1\over(\lambda\epsilon)^2} + \mathcal{O}\left({e^{-\epsilon \lambda}\over  (\lambda\epsilon)^3} \right).
\end{equation}
 There is no sign of any phase transition, and as $\lambda\epsilon\to \infty$ becomes arbitrarily smooth. 
 Again it is very easy to get arbitrarily large universes, both in spatial and temporal extent.

\subsection{Summary} \label{sec: summary1}

For both the uniform and DeWitt measure, the path integral seems to be well behaved in $\lambda$, meaning that it does not exhibit critical behavior. (There do not seem to be any phase transitions.) The non existence of a critical point means that this model does not seem to have the usual continuum limit based on a 2$^{nd}$ order phase transition. The non-criticality of the pure 2D gravity QRC path integral was shown in \cite{Hamber:1985gw}, however once a higher derivative $R^2$ term was added to the action and the partition function was restricted to fixed area critical behavior was observed numerically \cite{Gross:1990fq}. Future numerical work might examine the analogous behavior for this Lorentzian model with an $R^2$ term in the action  (see \cite{Hamber:1984kx, Ambjorn:1992aw} for implementations in QRC and DT).  As we mention in the discussion, there are reasons to believe the absence of a phase transition might actually be beneficial. 

\section{Loop-to-Loop Amplitudes} \label{sec: loopamp}
\begin{figure} [htbp]
\centering
\begin{tikzpicture}[scale=0.75]
\draw (-3,-3) arc (180:360:3cm and 1cm);
\draw[dashed] (3,-3) arc (0:180:3cm and 1cm);
\draw (0,3)  ellipse (2cm and 1cm);
\draw (-3,-3) .. controls (-4,-1) and (-1.5,1) .. (-2,3);  
\draw (3,-3) .. controls (4,-1) and (1.5,1) .. (2,3);
\draw[dashed] (0,-3) -- (0,3) (0,-3) -- (-3,-3) (0,3)--(-2,3); 
\filldraw [black] (0,-3) circle(1pt) (0,3) circle (1pt);
\draw (0,-4) node[anchor=south] {L} (0,4) node[anchor=south] {K} (0,0) node[anchor=west] {T};
\end{tikzpicture}
\caption{A Loop-to-Loop Geometry}
\label{fig:twoloop}
\end{figure}
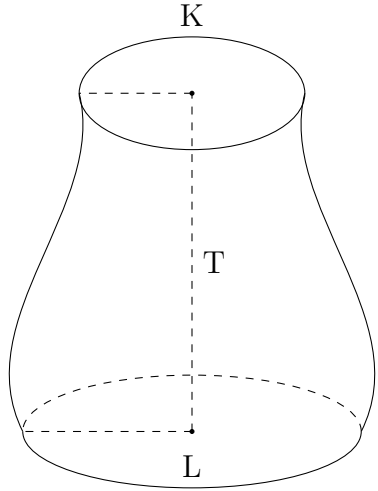
\subsection{$N_t$-step Amplitude} \label{sec: nstepamp}
An important quantitiy we would like to be able to calculate in the model is the loop-to-loop amplitude, $G_{\lambda}(L,K;T)$, which is the amplitude for a loop of length $L$ to propagate to a loop of length $K$ in ``time'' $T$. It will be a sum over all interpolating geometries of the type shown in figure \ref{fig:twoloop}. This amplitude has been calculated for CDTs in \cite{Ambjorn:1998xu} and in a continuum calculation in proper-time gauge in \cite{Nakayama:1993we}.

Unlike CDTs, in the current model there is no natural way to define the $T$ appearing in the loop-to-loop amplitude $G_{\lambda}(L,K;T)$. There is however a natural way to use the fixed triangulation to define the ``$N_t$-step'' loop-to-loop amplitude:
\begin{equation}
\label{eq: intloopamp}
G_{\lambda}(L,K; N_t).
\end{equation}

\begin{figure}[!htb]
\centering
\begin{tikzpicture}[scale=1.5]
\draw[step=1 cm,black,thick]
 (-5,1)  grid (4,2) ;
 \foreach \x in {-5,-4,...,3}
 	\draw[black,thick] (\x ,1 ) -- (\x+1 ,2 );
\draw (-5,1) node[anchor = east] {$L_i$} (-5,2) node[anchor=east] {$L_{i+1}$};
\draw (-3,1.5) node[anchor= north east] {$\bar{t}_n$} (-2.5,1.5) node[anchor=south east] {$t_n$};
\draw (-3.5,1) node[anchor=north]{$l_i$} (-2.5,2) node[anchor = south] {$l_{i+1}$};
 \end{tikzpicture}
 \caption{A Triangulated Strip}
 \label{fig: strip}
\end{figure}
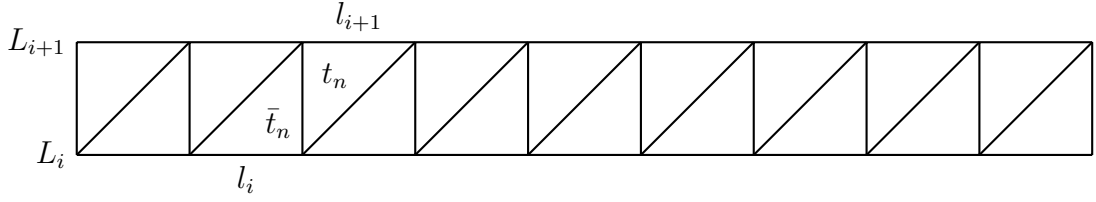

This amplitude will be given by equation \eqref{eq: Zwick} for a triangulated cylinder, where the spatial boundary lengths are fixed to be $L$ and $K$. For an $N_t \times N_s$ triangulation of the cylinder in the form of figure \ref{fig: tricyl}, let the space-like slices have length $L_i$ with $i=0,1,...,N_s$. The amplitude \eqref{eq: intloopamp} (for the uniform measure) is then given by:
\begin{equation}
\label{eq: intloopampmatrix}
G_{\lambda}(L,K; N_t) = \int \prod_{i=1}^{N_t-1} dL_i^2 \; G_{\lambda}(L,L_1;1)\; G_{\lambda}(L_1,L_2;1)\;\hdots\; G_{\lambda}(L_{N_t-1},K;1),
\end{equation}
where $G_{\lambda}(L_i,L_{i+1};1)$ is acting as a single-step temporal transfer matrix which we will call the strip amplitude. To construct the strip amplitude, consider the triangulated strip shown in figure \ref{fig: strip}. The space-like edges are fixed to be $l_i = L_i/N_s$ and $l_{i+1} = L_{i+1}/N_s$, however the vertical time-like edges labelled $\bar{t}_n$, and the diagonal time-like edges labelled $t_n$ with $n=1,...,N_s$, are dynamical degrees of freedom. 

The strip amplitude is then given by:
\begin{equation}
\label{eq: explicstripamp}
G_{\lambda}(L_i,L_{i+1};1)= \int_{0}^{\infty} \prod_{n=1}^{N_s} dt_n^2 \; d\bar{t}_n^2 \; e^{-\lambda \sum_{n=1}^{N_s} A_L(l_i,\bar{t}_n,t_{n-1}) + A_L(l_{i+1},\bar{t}_n,t_n)}.
\end{equation}
This can now be rewritten in terms of single-step spatial transfer matrices, $U$ and $V$, by making the identification $x_{2n-1} = \bar{t}^2_n$, $x_{2n} = t^2_{n}$, under which equation \eqref{eq: explicstripamp} becomes
\begin{eqnarray}
G_{\lambda}(L_i,L_{i+1};1) &=& \int_{0}^{\infty} dx_1 dx_2 \hdots dx_{2N_s} \nonumber\\
&&\qquad \times U(x_1,x_2)\;V(x_2,x_3)\hdots U(x_{2N_s-1},x_{2N_s})\;V(x_{2N_s},x_{1}) \nonumber \\
	&=& \textrm{Tr}[(UV)^{N_s}], \label{eq: intTS} 
\end{eqnarray}
where
\begin{eqnarray}
U(x_n,x_m) &=& e^{-\lambda A_L(l_{i+1},x_n,x_m)}, \label{eq: Tmat} \\
V(x_n,x_m) &=& e^{-\lambda A_L(l_i,x_n,x_m)}. \label{eq: Smat}
\end{eqnarray}
Thus, finding the strip amplitude is reduced to finding the spectrum (in fact since we can take $N_s$ to be large, the largest eigenvalue) of the integral operator
\begin{equation}
\label{eq: TS}
UV(x,y) = \int_{0}^{\infty} dz~ e^{-\frac{\lambda}{4} \left( \sqrt{(z+a)^2 + b^2} + \sqrt{(z+c)^2 +d^2} \right) }~.
\end{equation}
Here
\begin{align}
a &= l_{i+1}^2 - x; &\quad c &= l_i^2 -y; \nonumber \\
b &= 2l_{i+1}\sqrt{x}; &\quad d &= 2l_i\sqrt{y}. \label{eq: abcdxy}
\end{align}
It is unclear if any explicit solution to equation \eqref{eq: TS} can be found in terms of special functions. However in the special case where $l_i=l_j$, the diagonal of $UV$ can be readily computed:
\begin{eqnarray}
UV(x,x) &=& b \int_{\sinh^{-1}(a/b)}^{\infty} e^{-\frac{\lambda b}{2} \cosh(t)} ~ \cosh(t) ~dt \nonumber \\
 &=& 2 l_i \sqrt{x} \; \textrm{K}_1 \left[ \lambda l_i \sqrt{x}~ ,~ \sinh^{-1} \left( \frac{l_i^2-x}{2l_i \sqrt{x}} \right) \right]. \label{eq: TS diag}
\end{eqnarray}
Here $\textrm{K}_{1}[x,y]$ is an incomplete modified Bessel function of the second kind \cite{Jones: 2007aa, Harris:2008aa}.

Despite the difficulty of the integral appearing in equation \eqref{eq: TS}, we can make progress by investigating the integral operators $U$ and $V$ with kernels defined in equations \eqref{eq: Tmat} and \eqref{eq: Smat} individually. Considered as operators on $L^2_{\mathbb{R}}(0,\infty)$, that is the space of all real square integrable functions on $[0,\infty)$, they are positive, symmetric and bounded. For purposes of the following calculation temporarily identify $l_i^2$, $l_{i+1}^2  \rightarrow l$, and $\lambda /4 \rightarrow \lambda$, and use the inequality
\begin{equation}
\label{eq: upperbound}
\sqrt{(l+x+y)^2-4xy} \leq l+x+y, 
\end{equation}
which implies
\begin{equation}
\label{eq: lowerbound}
 e^{-\lambda \sqrt{(l+x+y)^2-4xy}} \geq e^{-\lambda(x+y+l)}.
\end{equation}
It is now easy to show that $U$ and $V$ are positive definite with respect to the $L^2$ inner product:
\begin{eqnarray}
(f,Uf) &=& \int_{0}^{\infty} dx \, f(x) \left( \int_{0}^{\infty} dy \, U(x,y) f(y) \right) \nonumber \\
          &=& \int_{0}^{\infty} \int_{0}^{\infty} dxdy \, e^{-\lambda \sqrt{(l+x+y)^2 - 4xy}} f(x)f(y) \nonumber \\
          &\geq& \int_{0}^{\infty} \int_{0}^{\infty} dxdy \, e^{-\lambda (l+x+y)} f(x)f(y) \nonumber \\
          &=& e^{-\lambda l} \left( \int_{0}^{\infty} dx \, e^{-\lambda x} f(x) \right)^2 > 0. \label{eq: posdef}
\end{eqnarray}
In addition the trace of $U$ and $V$ are finite, so the operators are trace class, and the traces are given by:
\begin{eqnarray}
\label{eq: Utrace}
Tr[U] &=& \int_{0}^{\infty} dx \, U(x,x) = \int_{0}^{\infty} dx \, e^{-\lambda \sqrt{l^2 + 4lx}}  
           = \frac{2(1+ \lambda l_i^2)}{\lambda^2 l_i^2} \; e^{-\frac{\lambda l_i^2}{4} },       
\end{eqnarray}
and
\begin{equation}
\label{eq: Vtrace}
Tr[V] = \frac{2(1+ \lambda l_{i+1}^2)}{\lambda^2 l_{i+1}^2} \; e^{-\frac{\lambda l_{i+1}^2}{4} }.
\end{equation}
These properties of $U$ and $V$ allow us to apply the Hilbert space generalization of the trace inequality \cite{Yang: 2000aa, Liu: 2007xx}:
\begin{equation}
\label{eq: traceineq}
Tr[(AB)^M] \leq (Tr[A])^M \, (Tr[B])^M.
\end{equation}
Thus, the strip amplitude is bounded above by:
\begin{eqnarray}
\label{eq: stripbound}
G_{\lambda}(L_i,L_{i+1};1) &=& Tr[(UV)^{N_s}] \nonumber \\
    &\leq& (Tr[U])^{N_s}\; (Tr[V])^{N_s} \nonumber \\
    &=&  4^{N_s} \; \frac{(1+ \lambda l_i^2)^{N_s} \; (1+ \lambda l_{i+1}^2)^{N_s}}{\lambda^{4N_s} \; l_i^{2N_s} \, l_{i+1}^{2N_s}} 
    \; e^{-\frac{\lambda N_s}{4}(l_i^2 + l_{i+1}^2) }.
\end{eqnarray}
Note that equation \eqref{eq: stripbound} is well behaved for all finite values of $\lambda$.

\subsection{Proper Time Amplitude} \label{sec: proptimeamp}
As we mentioned above, there is no natural ``proper time'' parameter appearing in the model. We can however introduce one, and the most obvious way to do so is to fix the ``vertical'' time-like edge lengths of the triangulation of figure \ref{fig: tricyl}  to a fixed common value $\tau=T/N_t$. This can be done by fixing all $\bar{t}_n=\tau$ in figure \ref{fig: strip}, while allowing the $t_n$ to remain dynamical. Once this is done, the flat space (Minkowski) solution for an $N_t \times N_s$ triangulated cylinder with time-like length (height) $T$ and space-like length (radius) $L$ is clearly $-t^2 = -\tau^2 + l^2 = - (T/N_t)^2 +(L/N_s)^2$, where $\tau = T/N_t$ is the time-like edge length of the vertical edges, $l = L/N_s$ is the space-like edge length of the horizontal edges, and $t$ is the time-like length of diagonal edges. Non-flat 1+1 geometries correspond to allowing $t$ to fluctuate from its Minkowski value. As above, to calculate the loop amplitude $G_{\lambda}(L,K;T)$ one must integrate over the diagonal edge lengths:
\begin{equation}
\label{eq: loopampT}
G_{\lambda}(L,K;T) = \int \prod_{i=1}^{N_t-1} dL_i^2 \; G_{\lambda}^{\tau}(L,L_1;1)\; G_{\lambda}^{\tau}(L_1,L_2;1)\; \hdots \; G_{\lambda}^{\tau}(L_{N_t-1},K;1),
\end{equation}
where now we need to find the strip amplitude with fixed ``vertical'' time-like edges $G_{\lambda}^{\tau}(L_i,L_{i+1};1)$. Notice that when the vertical edge lengths are fixed, the dynamical edge lengths only couple two neighboring triangles together, thus the strip amplitude factorizes:
\begin{equation}
\label{eq: stripfactor}
G_{\lambda}^{\tau} (L_i,L_{i+1};1) =  \left( \int_{0}^{\infty} dx \; e^{-\lambda A(x;l_i^2,l_{i+1}^2,\tau^2)} \right)^{N_s},
\end{equation}
where
\begin{equation}
\label{eq: Afunction}
A(x;a,b,c) = \frac14 \left( \sqrt{x^2 + 2(a-c)x + (a+c)^2} + \sqrt{x^2 + 2(b-c)x + (b+c)^2} ~ \right).
\end{equation}
Again, the integral found in equation \eqref{eq: stripfactor} is of the same form as that in equation \eqref{eq: TS} and whether a closed form solution can be found is at this stage unclear. However, the diagonal elements of the strip amplitude can be found by solving the integral in equation \eqref{eq: stripfactor} with $L_i = L_{i+1} = L$:
\begin{equation}
\label{eq: stripampdiag}
G_{\lambda}^{\tau} (L,L;1) = \left( \frac{2LT}{N_t N_s} \; \textrm{K}_1\left[ \frac{\lambda L T}{N_t N_s},\; \sinh^{-1}\left( \frac{(N_s/N_t)L^2-(N_t/N_s)T^2}{2LT} \right) \right] \right)^{N_s}.
\end{equation}
The strip amplitude can also be estimated for large $\lambda$ by performing a saddle point approximation. In this case, with $N_t=N_s$ for simplicity, the strip amplitude is approximated by:
\begin{equation}
\label{eq: saddlepoint}
G_{\lambda}(L_i,L_{i+1};1) = 
\left \{ 
\begin{array}{l l}  \left(\frac{2 \pi}{\lambda N_s^2} \, \frac{L_iL_{i+1}}{T^2} \, \frac{\left((L_i-L_{i+1})^2+4T^2\right)^{\frac32}}{L_i+L_{i+1}}  \right)^{N_s/2} 
\\  
\quad \quad \vphantom{\Bigg|} \times e^{-\frac{\lambda}{4N_s} \sqrt{ (L_i-L_{i+1})^2 +4T^2}~(L_{i} + L_{i+1})} & \quad L_iL_{i+1} < T^2;
\\  \\ 
\left( \frac{1}{2\lambda} \frac{(L_i^2+T^2)(L_{i+1}^2+T^2)}{L_i^2L_{i+1}^2-T^4} \right)^{N_s} 
\\ 
\quad \quad \vphantom{\Bigg|} \times  e^{-\frac{\lambda}{4N_s}(L_i^2+L_{i+1}^2+2T^2 )} & \quad L_iL_{i+1} > T^2. \\ 
\end{array} \right.
\end{equation}
%
\subsection{Summary} \label{sec: loopampsum}
The expressions derived above can be compared with the analgous expression in CDT \cite{Ambjorn:1998xu}. The most important fact is that the CDT expression has a non-zero critical value for $\lambda$,  while we have shown that equation \eqref{eq: intTS} is bounded above by an expression which is well behaved for all finite $\lambda$. It should not be unexpected that the quantities calculated in the different theories should be different, perhaps radically different, at the discrete level. However, it is clear that while in CDT a continuum limit can be taken by looking at a scaling limit near the critical value of $\lambda$, this does not seem to be possible for the current Lorentzian model. The only sensible way to compare the two theories is in their continuum limits, thus it is important that some way of taking a continuum limit be found for this model. As discussed in section \ref{sec: 1p1analysis} a possible way to do this would be to introduce an $R^2$ term into the action. That will be the topic of further research.

\section{The non-occurrence of spikes in LFT} \label{sec: spikes}
An problematic issue with Euclidean QRC, raised in reference \cite{Ambjorn:1997ub}, is what can be called the ``occurrence of spikes'' --- wherein ``spiky'' geometries dominate the path integral. Specifically, in (Euclidean) QRC for a fixed value of space-time area $A$ there exists a finite $n$ such that for any edge length in the triangulation
\begin{equation}
\label{eq: qrcspike}
\langle l^n \rangle_{A} = \infty.
\end{equation}
We will now argue that in the full LFT model these spikes \emph{do not} appear, and we will show by explicit calculation that they also \emph{do not} appear in the Causal Diamond truncation.

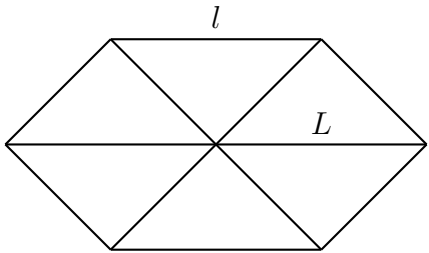
\begin{figure}[htbp]
\begin{center}
\begin{tikzpicture}[scale=1.4]
\def \n{2}
\def \s{0.5}
\def \d{1.5}
\draw[thick] (-2,0)--(2,0);
\draw[thick] (-1,1)--(1,1);
\draw[thick] (-1,-1)--(1,-1);
\draw[thick] (-1,1)--(1,-1);
\draw[thick] (1,1)--(-1,-1);
\draw[thick] (-1,1)--(-2,0);
\draw[thick] (-1,-1)--(-2,0);
\draw[thick] (1,1)--(2,0);
\draw[thick] (1,-1)--(2,0);
\draw (1,0) node[anchor = south] {$L$};
\draw (0,1) node[anchor = south] {$l$};
\end{tikzpicture}
\caption{A hexagonal neighborhood of a vertex in the triangulation. In the Euclidean case spikes appear at the central vertex when all of the incident edge lengths become large $L$, while all of the boundary edge lengths become small, $l = \frac{1}{L}$ in order to keep the total area finite. This situation cannot occur in the Lorentzian case.}
\label{fig: hexagon}
\end{center}
\end{figure}

Because the triangulation introduced for our model is hexagonal (see figure \ref{fig: hexagon}), we will focus on the analogous geometry in reference \cite{Ambjorn:1997ub}. In the Euclidean model it is demonstrated that spikes appear by considering regions of the configuration space in which all of the incident edge lengths on some vertex become very large $\sim L$. To keep the total area of the triangulation fixed all of the edge lengths of the boundary hexagon become correspondingly very small $\sim l = 1/L$. Due to the form of the Euclidean area function \eqref{eq: HerArea}, one can take the limit $L \rightarrow \infty$ and keep the total area of the triangulation fixed since the area of each triangle in the hexagon scales like
\begin{equation}
\label{eq; eucscale}
A_E \sim lL = 1.
\end{equation}
However, this cannot occur in the Lorentzian model because due to the form of the Lorentzian area function \eqref{eq: triArea} if one tries to take the same limit the now positive quartic terms dominate in four of the hexagon's six triangles:
\begin{equation}
A_L \sim L^2 \rightarrow \infty.
\end{equation}
In two of the hexagon's triangles (the ones with both time-like edges $\sim L$) the quartic terms can be arranged to cancel and the areas remain bounded, however the total area of the hexagon must diverge. In fact,  just one of the triangle areas diverging is enough to show that the hexagonal spikes that occur in Euclidean QRC cannot occur the LFT model.

\begin{figure} [htbp]
\centering
\begin{tikzpicture}[scale=0.75]
\draw (-3,-5) arc (180:360:3cm and 1cm);
\draw[dashed] (3,-5) arc (0:180:3cm and 1cm);
\draw (0,5)  ellipse (2cm and 1cm);
\draw (-3,-5) .. controls (-3,-3) and (0,-2).. (0,-1);  
\draw (3,-5) .. controls (3,-3) and (0,-2) .. (0,-1);
\draw (0,1) .. controls (-0.5,1.5) and (-2,3) .. (-2,5);  
\draw (0,1) .. controls (0.5,1.5) and (2,3)  .. (2,5);
\draw[very thick] (0,-1.138)--(0,1.025);
\end{tikzpicture}
\caption{A Pinched Geometry. Compare with the ``A'' phase of CDT.}
\label{fig:pinched}
\end{figure}
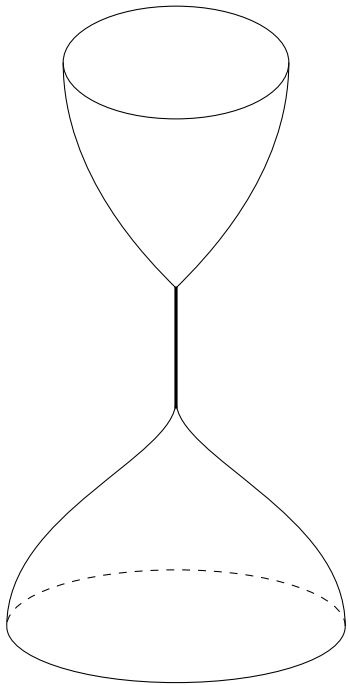

In fact, due to the form of \eqref{eq: triArea} the space-like edge lengths $S$ of the triangulation can never go beyond a finite bound if the area is to remain fixed.  As alluded to above, the only way one can scale the edge lengths of a Lorentzian triangle such that some edge lengths diverge while the total area remains fixed is to set $T_1=T_2 =L$ and $S=1/L$, in which case the area scales like
\begin{equation}
\label{eq: lorscale}
A_L \sim 1.
\end{equation}
Note however that for the triangulation we have considered in this model, if we attempt to scale a triangle in this way then the areas of the adjacent triangles, that is,  the ones to the left and right in the strip (see figure \ref{fig: strip}), will diverge unless they are scaled in an identical manner. This argument then extends to every triangle in the strip which contains the original triangle. Thus these regions in configuration space correspond to ``pinched'' geometries, that is geometries in which the spatial extent of a strip shrinks to zero while the temporal extent diverges (see figure \ref{fig:pinched}).  

The extent to which these pinched geometries will contribute to the partition function will depend on the how the parameters of the theory are tuned. We note that geometries very similar to this are dominant in the so-called ``\!$A$'' phase of CDT \cite{Ambjorn:2010ce}. It is possible then, perhaps only after adding an $R^2$ term to the action, that for fixed space-time area the general LFT model has phases where the pinched geometries are dominant and others where they are not. 

At any rate, these pinched geometries can be automatically regulated away by going to one of the two truncations introduced in the previous sections. In the proper time truncation,  the pinched geometries clearly cannot appear since half of the time-like edges in the triangulation are by definition frozen to a fixed value. This means that the scaling procedure outlined above cannot be performed. Perhaps more interesting is the Causal Diamond truncation in which the quantity $\langle l^n \rangle_{A}$ can be calculated exactly.

In the Causal Diamond truncation with fixed space-time area we need to consider the partition function 
\begin{equation}
\label{eq: ZcdfixedA}
Z[A] = \int \prod_{i} dx_i \; e^{\frac{\lambda}{2} \sum_{i} x_i} \; \delta \left( \frac12 \sum_{i} x_i - A \right) = 2 \; e^{-\lambda A} \; V_{M-1} = \frac{2^{M}}{(M-1)!} \; A^{M-1} \; e^{-\lambda A},
\end{equation}
where $M= N/2 = N_s N_t$ is the number of space-like edge lengths and $V_{M-1}$ is the volume of a right-angled $M-1$ simplex (in configuration space) which has all of its right-angle edge lengths equal to $2A$. Then
\begin{eqnarray*}
Z \,\langle l^n \rangle_A &=& 2 \, e^{-\lambda A} \int_{0}^{2A} dx_1 \cdots \int_{0}^{2A - \sum_{i=1}^{j-1} x_i} dx_j \cdots \int_{0}^{2A - \sum_{i=1}^{M-2}x_i} dx_{M-1} \; (x_{M-1})^{\frac{n}{2}} \\
				 &=& 2 \, e^{-\lambda A} \frac{1}{n/2 + 1} \int_{0}^{2A} dx_1 \cdots \int_{0}^{2A - \sum_{i=1}^{M-3}x_i} dx_{M-2} \; \left(2A - \sum_{i=1}^{M-3}x_i - x_{M-2} \right)^{\frac{n}{2}+1} \\
				 &=& 2 \, e^{-\lambda A} \frac{1}{\frac12( n + 2)} \int_{0}^{2A} dx_1 \cdots \int_{0}^{2A - \sum_{i=1}^{M-3}x_i} d\bar{x} \; \left( \bar{x} \right)^{\frac{n}{2}+1} \\
				 &\vdots& \\
				 &=& \frac{ 2^{2M-1+\frac{n}{2}} \, A^{M-1 + \frac{n}{2}} e^{-\lambda A} }{(n+2)(n+4) \cdots(n+2(M-1))}.		 
\end{eqnarray*}
One performs the iterated integration above by repeatedly making the linear change of variables $\bar{x} = 2A - \sum_{i}^{j-1}x_i - x_j$. Thus the expectation value is given by:
\begin{equation}
\label{eq: lnCD}
\langle l^n \rangle_A = 2^{\frac{n}{2} -1}\; n \; \frac{ \Gamma \left(M \right) \Gamma \left(\frac{n}{2}\right)}{\Gamma \left( M + \frac{n}{2} \right)} \; A^{\frac{n}{2}}.
\end{equation}
This is finite for any finite $n$ and thus the appearance of spikes is completely removed from the Causal Diamond model. A quick sanity check with $n=2$ reveals:
\begin{equation}
\frac{\langle l^2 \rangle_A}{2} = \frac{A}{M},
\end{equation}
exactly as should be expected for a fixed area. Note also that
\begin{equation}
\langle l^n \rangle_A \sim A^{\frac{n}{2}} \sim \langle l \rangle_A^n,
\end{equation}
so a variant of the scaling argument presented in the discussion following equation (37) of reference~\cite{Ambjorn:1997ub} implies that  the fractal dimension for the Causal Diamond model is simply its topological dimension, namely 2.
We remark that in the ``thermodynamic limit'', where $M\to \infty$ holding $n$ and $A$ fixed, we have the stronger result 
\begin{equation}
\langle l^n\rangle_A =    \Gamma(1+{\textstyle{n\over2}}) \; \langle l^2 \rangle_A^{n/2},
\end{equation}
while the average area per triangle in this limit is simply ${1\over2}\langle l^2\rangle_A$. 



In view of these comments, it is therefore clear that (at least in two particularly interesting truncations of the LFT model) the spikes which plague the Euclidean QRC simply do not occur. We have also argued that the QRC spikes cannot occur in the full model either, however there are still regions of the configuration space which correspond to ``pinched'' geometries. Whether these geometries will dominate the partition function (and if they do whether or not that would be pathological or interesting) will be the topic of further research.

\begin{figure}[htb]
\centering
\begin{minipage}[b]{5 cm}
\begin{tikzpicture}[scale=4.0]
\draw (0.2,0.2) -- (-0.4,0.45);
\draw (0.2,0.2) -- (0.5,0.55);
\draw (0.2,0.2) -- (0.15,1.2);
\draw (-0.4,0.45) -- (0.15,1.2);
\draw (0.5,0.55) -- (0.15,1.2);
\draw[dashed]  (-0.4,0.45)--(0.5,0.55);
\draw (-0.1,0.325) node [anchor= north east] {$S_1$};
\draw (0.375,0.390) node [anchor = north west] {$S_2$};
\draw (-0.05,0.5) node [anchor= south ] {$S_3$};
\draw (0.175,0.7) node[anchor = south east] {$T_3$};
\draw (-0.125,0.825) node [anchor = south east ] {$T_2$};
\draw (0.325,0.875) node [anchor = west] {$T_1$};
\end{tikzpicture}
\caption{(3,1) tetrahedron}
\label{fig: 31tet}
\end{minipage}
\hfil
\begin{minipage}[b]{5cm}
\begin{tikzpicture}[scale=4.25]
\draw (0,0)--(-0.65,0.05);
\draw (0,0)--(-0.4,0.6);
\draw (-0.65,0.05)--(-0.4,0.6);
\draw (0,0) -- (0.1,0.7);
\draw (-0.4,0.6)--(0.1,0.7);
\draw[dashed] (-0.65,0.05) -- (0.1,0.7);
\draw (-0.325,0.025) node [anchor = north] {$S_1$};
\draw (-0.2,0.3) node[anchor = west] {$T_2$};
\draw (-0.525,0.3025) node [anchor = east] {$T_1$};
\draw (0.05,0.350) node [anchor = west] {$T_3$};
\draw (-0.150,0.650) node [anchor = south] {$S_2$};
\draw (-0.4,0.3) node [anchor = north west ] {$T_4$};
\end{tikzpicture}
\caption{(2,2) tetrahedron}
\label{fig: 22tet}
\end{minipage}
\end{figure}

\section{Higher Dimensions} \label{sec: highdim}

The model discussed in the previous sections can be extended into higher dimensions with a few modifications. The first is that in $n$ dimensions we now need to consider the Einstein--Hilbert term in the gravitational action \cite{Hamber:1984tm}:
\begin{equation}
\label{eq: highdimS}
S = \frac{k}{2} \int d^nx~\sqrt{g} (R-2\Lambda) \rightarrow \kappa \sum_{h} V_h \delta_{h} - \lambda \sum_{s} V_L(s),
\end{equation}
where $h$ denotes the $n-2$ dimensional ``hinges'' in the triangulation along which curvature is concentrated, $V_h$ is the Lorentzian volume of a hinge, $\delta_h$ is the deficit angle of the hinge and $V_L(s)$ is the Lorentzian volume of the $n$-dimensional simplices $s$. 

This model will differ from QRC in three ways: first, the volumes of the simplices will be determined by their Lorentzian formulas, which can be derived from the analogous Euclidean formulas in a way similar to equation \eqref{eq: Area-Lor-Eu}; the exact procedure is outlined in the next section. Second, the nature of the deficit angle of a hinge will depend on whether that hinge is space-like, or time-like, or perhaps even null (see \cite{Sorkin:1975ah}). Finally, there will be generalized triangle inequality constraints on the edge length configuration space, however these will \emph{not} be the constraints found in the Euclidean theory as there will be certain simplices in the triangulation whose edge-lengths are constrained and others whose edge-lengths are not; we will derive the Lorentzian generalized triangle inequalities in $2+1$ dimensions below. The Euclidean and Lorentzian QRC models will be related in a manner similar to equation \eqref{eq: QRCLFTrelate}, which is clearly much more involved then a simple, naive Wick rotation. In light of this, studying the Lorentzian QRC calculus numerically will involve adapting previous studies of Euclidean QRC by implementing the changes described above. Analytic investigations will be mainly constrained by the same obstacles encountered in Euclidean QRC, and will therefore by strongly limited. 

In developing the model in 2+1 dimensions it will, in general, be impossible to triangulate the manifold in such a way that every face of the Lorentzian tetrahedron can be taken to be a Lorentzian triangle. There are 4 types of Lorentzian tetrahedron; two of them are very simply related to the Euclidean tetrahedron: one with all six edges space-like, i.e., the standard Euclidean tetrahedron itself with the usual Euclidean triangle inequalities, and the trivial generalization in which all six edge lengths are time-like, in this case all triangle inequalities are reversed (due to the $-1$ in the metric) but the tetrahedral conditions are unchanged. 

As described in \cite{Ambjorn:2001cv}, for the types of manifolds we are interested in, there are two types of tetrahedra which need to be considered: the (3,1) type in which a Euclidean triangle in one time slice is connected to a vertex in a subsequent time slice by three time-like edges (see figure \ref{fig: 31tet}); the other is the (2,2) type in which there are two space-like edges, each in a separate time slice, connected by four time-like edges (see figure \ref{fig: 22tet}). As we will soon demonstrate the (3,1) tetrahedron is  analogous to the Lorentzian triangle in that its time-like edge lengths are completely unconstrained (its space-like edge lengths form a Euclidean triangle and therefore must satisfy the Euclidean triangle inequalities); however, all edge lengths of the (2,2) tetrahedron must satisfy complicated constraints.

\subsection{(3,1) Lorentzian Tetrahedron} \label{sec: 31lortet}
The question is, under what conditions is the sextuplet $S_{(3,1)} = \left\{S_1,S_2,S_3,T_1,T_2,T_3\right\}$, where $S_i$ is opposite $T_i$,  a (3,1) tetrahedron? First the tetrahedron has four faces, one is a standard Euclidean triangle with edge lengths $S_1$, $S_2$ and $S_3$, and the remaining three are Lorentzian triangles with one space-like edge and two time-like edges. Thus the only requirement on the faces is that the edges of the space-like triangle satisfy the Euclidean triangle inequalities:
\begin{eqnarray}
S_1 &<& S_2 + S_3, \nonumber \\
S_2 &<& S_3 + S_1, \nonumber \\
S_3 &<& S_1 + S_2. \label{eq: triineq}
\end{eqnarray}
Note that having the faces satisfy these conditions  only means that the 4 triangles formed by $S_{(3,1)}$ are at this stage merely a ``facial net'' in the language of \cite{Wirth:2009aa}. To form a Lorentzian tetrahedron the volume must be a positive real number, equivalently the  Cayley--Menger determinant must be positive. 

The volume of a Euclidean $n$-simplex is given by:
\begin{equation}
\label{eq: EucCMdet}
\left(V_n\right)_E^2 = \frac{(-1)^{n+1}}{2^n (n!)^2} |E_{ij}^2|,
\end{equation}
where $E_{ij}^2$ is the $(n+2) \times (n+2)$ matrix of  squared Euclidean distances between the $n+1$ vertices $i$ and $j$ in the simplex, augmented with an additional row and column defined by $E_{00}^2 = 0$, $E_{i0}^2 = 1$ and $E_{0j}^2 = 1$. The determinant of this matrix is known as the Cayley--Menger determinant. For example, a triangle with side lengths $a$, $b$, and $c$,  has area given by:
\begin{equation}
\label{eq: HerCMdet}
(A_2)_E^2 = -\frac{1}{16} \left|
\begin{array}{cccc}
    0 & 1 & 1 & 1 \\
    1 & 0 & a^2 & b^2 \\
    1 & a^2 & 0 & c^2  \\
    1 & b^2 & c^2 & 0  
\end{array} \right| ,
\end{equation}
which results in Heron's formula \eqref{eq: HerArea}. 

Note that this determinant is positive \textit{iff} (if and only if) $a$, $b$ and $c$ satisfy the triangle inequalities. To find the analogous formula for Lorentzian simplices we make use of the following observation. Given a Euclidean simplex with edge lengths $\{L_i\}$, if we partition this set into two sets $\{S_i\}$ and \{$T_i$\} then the analogous Lorentzian simplex, with time-like edge lengths $\{T_i\}$ and space-like edge lengths $\{S_i\}$, has volume given by:
\begin{equation}
\label{eq: LorEucVolRelate}
V_L[\{T_i\},\{S_i\}] = -i V_E[\{iT_i\},\{S_i\}].
\end{equation}
Thus, $V_L^2 = -V_E^2$, and therefore the volume of a Lorentzian $n$-simplex is given by:
\begin{equation}
\label{eq: LorCMdet}
\left(V_n\right)_L^2 = \frac{(-1)^{n}}{2^n (n!)^2} |L_{ij}^2|,
\end{equation}
where $L_{ij}^2$ is the now Lorentzian distance squared from vertex $i$ to $j$. Thus the area of a Lorentzian triangle with space-like edge length $S$ and time-like edge lengths $T_1$ and $T_2$ is given by
\begin{equation}
\label{eq: LorTriCMdet}
(A_2)_L^2 = \frac{1}{16} \left|
\begin{array}{cccc}
    0 & 1 & 1 & 1 \\
    1 & 0 & S^2 & -T_1^2 \\
    1 & S^2 & 0 & -T_2^2  \\
    1 & -T_1^2 & -T_2^2 & 0  
\end{array} \right| ,
\end{equation}
which results in the formula we derived before, equation \eqref{eq: triArea}. Note that this determinant is positive for any choice of $S$, $T_1$, $T_2$. 

Finally, we come to the volume formula for the $(3,1)$ tetrahedron with edge lengths as above:
\begin{equation}
\label{eq: Vol31tet}
(V_3)_{(3,1)}^2 = -\frac{1}{288} \left|
\begin{array}{ccccc}
    0 & 1 & 1 & 1 & 1\\
    1 & 0 & S_1^2 & S_2^2 & -T_3^2 \\
    1 & S_1^2 & 0 & S_3^2 & -T_2^2 \\
    1 & S_2^2 & S_3^2 & 0 & -T_1^2 \\
    1 & -T_3^2 & -T_2^2 & -T_1^2 & 0
\end{array} \right| .
\end{equation}
Remarkably \eqref{eq: Vol31tet} is positive for arbitrary real values of $T_1$, $T_2$, and $T_3$, provided that \eqref{eq: triineq} is satisfied. 
That is, one need only apply triangle inequalities to the single Euclidean face, the three Lorentzian faces are unconstrained. 
A symmetric way of writing out the volume is:
\begin{eqnarray}
\textrm{Vol}(3,1) &=& \frac{1}{12} \bigg[     S_1^2T_1^2 \left( S_2^2+S_3^2-S_1^2 +T_1^2-T_2^2-T_3^2 \right) \nonumber \\
	                                                      &+&  S_2^2T_2^2 \left( S_3^2+S_1^2-S_2^2 +T_2^2-T_3^2-T_1^2 \right) \nonumber \\
	                                                      &+&  S_3^2T_3^2 \left( S_1^2+S_2^2-S_3^2 +T_3^2-T_1^2-T_2^2 \right)  \nonumber \\
	                                                      &+& \frac12 (S_1^2+T_1^2)(S_2^2+T_2^2)(S_3^2+T_3^2) +\frac12 (S_1^2-T_1^2)(S_2^2-T_2^2)(S_3^2-T_3^2) \bigg]^{1/2}~. \label{eq: form31vol1}
\end{eqnarray}
Another way to write out the volume, with $x = {1\over2}(S_2+S_3-S_1)$, $y={1\over2}(S_3+S_1-S_2)$, $z={1\over2}(S_1+S_2-S_3)$, and $s = {1\over2}(S_1+S_2+S_3) = x+y+z$, is:
\begin{eqnarray}
\textrm{Vol}(3,1) &=& \frac{1}{12} \bigg[ (s-x)^2 T_1^4 + (s-y)^2 T_2^4 + (s-z)^2 T_3^4 \nonumber \\
                                                              &+& 2(s-x)^2T_1^2(sx-yz) +2(s-y)^2T_2^2(sy-xz) + 2(s-z)^2T_3^2(sz-xy) \nonumber \\
                                                              &-& 2T_1^2T_2^2(sz-xy) - 2T_1^2T_3^3(sy-xz) - 2T_2^2T_3^2(sx-yz)  \nonumber \\
                                                              &+& (s-x)^2(s-y)^2(s-z)^2 \bigg]^{1/2}~. \label{eq: form31vol2}
\end{eqnarray}

The formula for the cosine of the dihedral angles is found by making the appropriate substitutions into the formula for a Euclidean tetrahedron. For the Euclidean case, with edge lengths $a$ opposite to $d$, $b$ opposite to $e$, and $c$ opposite to $f$, and the four faces given by $\triangle abf$, $\triangle bcd$, $\triangle ace$, and $\triangle def$, the cosine of the dihedral angle at the edge $a$ is given by:
\begin{equation}
\label{eq: cosdihedEuc}
\cos \theta_{a} = \frac{ \left[ -a^4 - (c^2-e^2)(b^2-f^2) + a^2 \left(b^2+c^2-2d^2+e^2+f^2 \right) \right]}{16\,A_{E}(a,c,e) \; A_{E}(a,b,f)}. 
\end{equation}
In this case then the cosine of the dihedral angle at a space-like edge $\cos \theta_{S}$ and the dihedral angle at a time-like edge $\cos \theta_{T}$ are given by:
\begin{eqnarray}
\cos \theta_{S_1} &=&  \frac{ \left[ -S_1^4 - (T_2^2-T_3^2)(S_2^2-S_3^2) + S_1^2 \left(S_2^2+S_3^2+2T_1^2-T_2^2-T_3^2 \right) \right]}{16i~A_L(S_1,T_2,T_3) \; A_E(S_1,S_2,S_3) },  \label{eq: cosdihedS} \\
\cos \theta_{T_1} &=&  \frac{ \left[ T_1^4 + (S_2^2+T_2^2)(S_3^2+T_3^2) - T_1^2 \left(T_2^2+T_3^2+2S_1^2-S_2^2-S_3^2 \right) \right]}{16\,A_L(S_2,T_1,T_3) \; A_L(S_3,T1,T2)}. \label{eq: cosdihedT}
\end{eqnarray}

The sine of the dihedral angle is much easier to find by using the Lorentzian analog of the Euclidean $d$-dimensional sine law:%
\begin{equation}
\label{eq: Eucsine}
\sin \theta_E = \frac{d}{d-1}\; \frac{V_{n} \; V_{n-2}}{V_{n-1} \; V'_{n-1}},
\end{equation}
where $V_n$ is the volume of the $n$-simplex, $V_{n-2}$ is the volume of the hinge at which the dihedral angle is being considered, and $V_{n-1}$ and $V'_{n-1}$ are the volumes of the two faces which share the hinge. From this we can derive the sine of the dihedral angle about a space-like edge length, $\sin \theta_{S}$, and about a time-like edge length, $\sin \theta_{T}$:
\begin{eqnarray}
\sin \theta_{S_1} &=& \frac{3 S_1}{2} \frac{V_{(3,1)}(S_1,S_2,S_3,T_1,T_2,T_3)}{A_L(S_1,T_2,T_3) \; A_E(S_1,S_2,S_3)}, \label{eq: sindihedS} \\
\sin \theta_{T_1} &=& \frac{3 T_1}{2} \frac{V_{(3,1)}(S_1,S_2,S_3,T_1,T_2,T_3)}{A_L(S_2,T_1,T_3) \; A_L(S_3,T_1,T_2)}. \label{eq: sindihedT} 
\end{eqnarray}
We note that equations \eqref{eq: form31vol2}, \eqref{eq: cosdihedS}, \eqref{eq: cosdihedT}, \eqref{eq: sindihedS}, and \eqref{eq: sindihedT}, reproduce the results found in \cite{Ambjorn:2001cv} for the special case $S_1=S_2=S_3=1$ and $T_1=T_2=T_3=\sqrt{\alpha}$.

\subsection{(2,2) Lorentzian Tetrahedron} \label{sec: 22lortet}
The sextuple of edge lengths for the (2,2) tetrahedron is $S_{(2,2)} = \{S_1,S_2,T_1,T_2,T_3,T_4\}$, where $S_1$ is opposite $S_2$, $T_1$ is opposite $T_3$ and $T_2$ is opposite $T_4$. The volume is then given by:
\begin{equation}
(V_3)_{(2,2)}^2 = -\frac{1}{288} \left|
\begin{array}{ccccc}
    0 & 1 & 1 & 1 & 1\\
    1 & 0 & S_1^2 & -T_1^2 & -T_4^2 \\
    1 & S_1^2 & 0 & -T_2^2 & -T_3^2 \\
    1 & -T_1^2 & -T_2^2 & 0 & S_2^2 \\
    1 & -T_4^2 & -T_3^2 & S_2^2 & 0
\end{array} \right| .
\end{equation}
The constraints on the edge lengths are now much more complicated than simple triangle inequalities. If the edge lengths are considered as vectors, then one can derive $\vec{T}_1 + (-\vec{T}_2) + \vec{T}_3 + (-\vec{T}_4)=0$. That is, starting at the $(S_1,T_1,T_4)$ vertex in figure \ref{fig: 22tet} one can follow the closed path defined by the previous equation. This implies that $\vec{T}_1 + \vec{T}_3 = \vec{T}_2 + \vec{T}_4$, which is equivalent to saying the four time-like edges form a non-planar quadrilateral, and are therefore constrained --- they cannot take on arbitrary values.\footnote{Though we can easily solve this particular constraint by setting $\vec{T}_{1,3} = \vec{a} \pm \vec{b}$ and $\vec{T}_{2,4} = \vec{a} \pm \vec{c}$, this merely moves the problem elsewhere. We then have to constrain $\vec{a}$, $\vec{b}$, $\vec{c}$ by demanding that $\vec{a} \pm \vec{b}$ and $\vec{a} \pm \vec{c}$ all be timelike.} An example of a sextuple which is an impossible (2,2) tetrahedron is $S_{(2,2)} = \{1/2,1,\sqrt{2}, 1/\sqrt{2},1,1\}$. A symmetrical representation of the volume formula is:
\begin{eqnarray}
\textrm{Vol}(2,2) &=& \frac{1}{12} \bigg[     S_1^2S_2^2 \left( S_1^2+S_2^2+T_1^2 +T_2^2+T_3^2+T_4^2 \right) \nonumber \\
	                                                      &&\qquad-     T_1^2T_3^2 \left( S_1^2+S_2^2+T_1^2 -T_2^2+T_3^2-T_4^2 \right) \nonumber \\
	                                                      &&\qquad-     T_2^2T_4^2 \left( S_1^2+S_2^2-T_1^2 +T_2^2-T_3^2+T_4^2 \right)  \nonumber \\
	                                                      &&\qquad+ \frac12 (S_1^2+S_2^2)(T_1^2+T_3^2)(T_2^2+T_4^2) \nonumber\\
	                                                      &&\qquad -\frac12 (S_1^2-S_2^2)(T_1^2-T_3^2)(T_2^2-T_4^2) \bigg]^{1/2}. \quad \label{eq: vol22}
\end{eqnarray}
It appears that within the framework outlined in this paper the (2,2) tetrahedron re-introduces the difficulty of configuration space constraints on the time-like edge lengths. This will impede analytic calculations in the same way the Euclidean constraints impede calculations in QRC. Therefore studying this model in higher dimensions will best be done numerically. 

In the same manner as in section \ref{sec: 31lortet} we can find the cosine of the dihedral angles at a space-like edge length, $\cos \theta_S$, and a time-like edge length, $\cos \theta_T$:
\begin{eqnarray}
\cos \theta_{S_1} &=&  \frac{ \left[ S_1^4 - (T_1^2-T_2^2)(T_3^2-T_4^2) + S_1^2 \left(T_1^2+T_2^2+T_3^2+T_3^2+2S_2^2 \right) \right]}{16\,A_L(S_1,T_1,T_2) \; A_L(S_1,T_3,T_4)}, \label{eq: cosdihed22S} \\
\cos \theta_{T_1} &=&  \frac{ \left[ T_1^4 - (S_1^2+T_2^2)(S_2^2+T_4^2) + T_1^2 \left(S_1^2+S_2^2+2T_3^2-T_2^2-T_4^2 \right) \right]}{16\,A_L(S_1,T_1,T_2) \; A_L(S_2,T_1,T_4)} \label{eq: cosdihed22T}. 
\end{eqnarray}
We can also find the sine of the dihedral angles at a space-like edge length, $\sin \theta_S$, and a time-like edge length, $\sin \theta_T$:
\begin{eqnarray}
\sin \theta_{S_1} &=& -i \frac{3 S_1}{2} \frac{ V_{(2,2)}(S_1,S_2,T_1,T_2,T_3,T_4) }{A_L(S_1,T_1,T_2) \; A_L(S_1,T_3,T_4)}, \label{eq: sindihed22S} \\
\sin \theta_{T_1} &=&  \frac{3 T_1}{2} \frac{ V_{(2,2)}(S_1,S_2,T_1,T_2,T_3,T_4) }{A_L(S_1,T_1,T_2) \; A_L(S_2,T_1,T_4)}. \label{eq: sindihed22T}
\end{eqnarray}
Equations \eqref{eq: vol22}--\eqref{eq: sindihed22T} also reduce to the values found in reference \cite{Ambjorn:2001cv} for the specific values $S_1=S_2=S_3=1$ and $T_1=T_2=T_3=\sqrt{\alpha}$.

\subsection{$3+1$ and Beyond} \label{sec: 3p1B}
The procedure carried out in sections \ref{sec: 31lortet} and \ref{sec: 22lortet} can be applied in exactly the same way in any dimension $n=d+1$, however with increasing difficulty as the number of edge lengths grows like $n^2$. In $3+1$ dimensions there are again two types of 4-simplex that need to be considered: the (4,1) type which is the analog of the (3,1) tetrahedron, and the (3,2) type which is the analog of the (2,2) tetrahedron \cite{Ambjorn:2001cv}. Finding the volume and the sine of the dihedral angles is simply a matter of applying equations \eqref{eq: LorCMdet} and \eqref{eq: cosdihedEuc} respectively. Finding the cosine of the dihedral angle is more involved but can be done using the higher dimensional analog of the cosine law. Explicit formulas for these will simply be long cumbersome analogs of those presented above. One question that can be asked is what are the generalized Lorentzian signature triangle inequalities for these Lorentzian simplices? One can show that the (4,1) simplex has its 4 time-like edge lengths completely unconstrained, provided its Euclidean tetrahedron is realizable, while the (3,2) simplex has its edge lengths subject to complicated constraints \cite{Tate:2011rm}. Thus, it is safe to say that the constrained Lorentzian configuration space is very different from that of the Euclidean configuration space in any dimension. 
    
\section{Conclusions} \label{sec: conc}
We have formulated a model of simplicial Quantum Gravity which is analgous to the Quantum Regge Calculus, but in the Lorentzian domain. This model is formulated by using an insight from the development of CDTs, which is to use the structure of Lorentzian manifolds to guide which geometries appear in the path integral. It is interesting to note that in CDTs this results in decreasing the set of allowed geometries to those which have a causal structure; however in our model we increase the number of geometries by modifying and relaxing the triangle inequalities. In 1+1 dimensions the removal of the triangle inequalities allows every dynamical edge length in the model to be integrated over in a completely unconstrained fashion. This allows more freedom for analytical calculation,  which we have used to derive scaling relations for the pure gravity theory. Unfortunately obstacles remain for both full calculation of the path integral and also for calculation of loop amplitudes. In calculating loop amplitudes, the integrals that arise have no known closed form solution and block us from deriving the full result. Unless explicit results for these integrals are found, further analytical work will have to proceed in different directions.

Another  aspect of this model is that, in contrast to CDT, the pure gravity path integral does not exhibit any critical behavior. This means that there is no way to take the usual sort of  continuum limit of the model which could be compared to other approaches. This same aspect is found in the Euclidean version of the model, and this is resolved in that model by introducing higher derivative terms into the action \cite{Hamber:1985gw}. The exact same procedure can be done in this model by using the angle formulas given in equations \eqref{eq: ST1angle}--\eqref{eq: T1T2angle}. This modified model will most likely need to be studied numerically through Monte Carlo simulations, as analytical work will run into the same obstacles found in the preceding sections. On the other hand, the lack of phase transition could be interpreted as a plus: A potential polymerized phase is excluded kinematically, while in 1+1 dimensions the occurrence of a rough/ fractal/ crumpled phase seems unlikely given the exact result $\langle R \rangle=0$ and the well-behaved nature of the causal diamond truncation. This strongly suggests the theory is always in a smooth phase. Large universes (with spatial size controlled by the number $N_s$ of triangles in the spatial direction, and temporal size controlled by the number $N_t$ of triangles in the temporal direction) seem to be the norm rather than the exception. 

There are prospects for further analytical work using the causal diamond truncation introduced in section \ref{sec: 1p1analysis}. Because in pure 1+1 gravity the only degree of freedom is the Liouville field, the causal diamond truncation could be justified since the varying space-like edge lengths act like a simplicial version of the Liouville field. Future work could focus on calculating quantities in this truncation such as the string susceptibility $\chi_{\phi}$.

The model is also safe from the occurrence of the spikes which trouble the Euclidean QRC. In both the proper-time and the Causal Diamond truncation of the model the partition function with fixed space-time area is well behaved with the appearance of spikes completely suppressed. In the full model, there are regions of the configuration space in which the geometry (while not spiky) can become ``pinched''. These regions may or may not be important, however if they are it is not clear that they are pathological; instead they may correspond to an interesting phase of the model, perhaps only controllable by adding an $R^2$ term in the action.

The model can be extended to higher dimensions, however generalized Lorentzian triangle inequalities are introduced, and difficulties analogous to the ones found in the Euclidean theory appear. The Lorentzian model in higher dimensions is not related to the Euclidean model by simple Wick rotation, as there are still sets of edge lengths which are unconstrained, and therefore the configuration space of the Lorentzian model is larger than that of the Euclidean one. The model in higher dimensions will need to investigated numerically, and it will be interesting to see how the results compare with those of CDT. 

In summary, in this paper we have demonstrated that Quantum Regge Calculus in the Lorentzian domain differs from that in the Euclidean domain in ways considerably more complicated than by a simple Wick rotation. This is a specific case of the general fact that Lorentzian and Euclidean quantum gravity are not trivially connected. In order to study Lorentzian quantum gravity using QRC it is therefore necessary to take into consideration the differences we have outlined in this paper.


\end{document}